\documentclass[aps,superscriptaddress,twocolumn,superscriptaddress,showpacs,showkeys,
groupedaddress,nofootinbib,nobalancelastpage,nobibnotes]{revtex4}
\pdfoutput=1
\usepackage{amsmath,amsfonts,amssymb,mathrsfs,graphicx,color,epsfig}
\usepackage[squaren]{SIunits}
\usepackage{verbatim}
\usepackage{enumerate}
\usepackage[hyperfootnotes=false]{hyperref}
\usepackage{xcolor}
\usepackage{slashed}
\usepackage[utf8]{inputenc}
\usepackage{rotating}
\usepackage{isotope}

\newcommand{\ie}{{\it i.e.}}

\newcommand{\eg}{{\it e.g.}}

\newcommand{\eq}{Eq.}

\newcommand{\fig}{Fig.}

\newcommand{\Ref}{Ref.}
\newcommand{\Refs}{Refs.}

\newcommand{\equ}[1]{\eq~(\ref{equ:#1})}
\newcommand{\figu}[1]{\fig~\ref{fig:#1}}

\newcommand{\bi}{\begin{itemize}}
\newcommand{\ei}{\end{itemize}}

\newcommand{\uh}{UHECR}
\newcommand{\td}{TDEs}
\newcommand{\ns}{neutrinos}
\newcommand{\n}{neutrino}
\newcommand{\sbh}{SMBH}
\newcommand{\wdw}{WD}
\newcommand{\bh}{IMBH}
\newcommand{\msun}{M_{\odot}}

\begin{document}

\title{Tidally disrupted stars as a possible origin of both cosmic rays and neutrinos \newline at the highest energies}

\author{Daniel Biehl}
\affiliation{Deutsches Elektronen-Synchrotron (DESY), Platanenallee 6, D-15738 Zeuthen, Germany}

\author{Denise Boncioli}
\affiliation{Deutsches Elektronen-Synchrotron (DESY), Platanenallee 6, D-15738 Zeuthen, Germany}

\author{Cecilia Lunardini}
\affiliation{Department of Physics, Arizona State University, 450 E. Tyler Mall, Tempe, AZ 85287-1504 USA}

\author{Walter Winter}
\affiliation{Deutsches Elektronen-Synchrotron (DESY), Platanenallee 6, D-15738 Zeuthen, Germany}

\date{\today}

\begin{abstract}
\vspace{0.5cm}
Tidal Disruption Events (TDEs) are processes where stars are torn apart by the strong gravitational force near to a massive or supermassive black hole. If a jet is launched in such a process, particle acceleration may take place in internal shocks. We demonstrate that jetted TDEs  can simultaneously describe the observed neutrino and cosmic ray fluxes at the highest energies if stars with heavier compositions, such as carbon-oxygen white dwarfs, are tidally disrupted and these events are sufficiently abundant. We simulate the photo-hadronic interactions both in the TDE jet and in the propagation through the extragalactic space and we show that the simultaneous description of Ultra-High Energy Cosmic Ray (UHECR) and PeV neutrino data implies that a nuclear cascade in the jet develops by photo-hadronic interactions. 
\end{abstract}

\maketitle

The discovery of high-energy ($\sim 0.1 -1$ PeV) astrophysical neutrinos~\cite{Aartsen:2013jdh} has triggered substantial research on their possible origin. These  neutrinos come probably from outside our galaxy, and can naturally arise from a flux of parent protons or nuclei.  These facts together with basic energy budget considerations~\cite{Waxman:1998yy} suggest that they may have the same origin as the Ultra-High Energy Cosmic Rays (UHECRs). At this time, however, a class of astrophysical objects that can be the common origin of both UHECRs and \ns\ has not been identified. Neutrino data analyses disfavor some of the traditional candidates, such as  Gamma-Ray Bursts (GRBs)~\cite{Aartsen:2017wea} and Blazars~\cite{Aartsen:2016lir}, and diffuse gamma-ray observations  already strongly constrain starburst galaxies~\cite{Murase:2013rfa,Bechtol:2015uqb}. Therefore the origin of the observed neutrinos remains a mystery, and alternatives are now being considered.

Tidal Disruption Events (\td) are one such alternative. Tidal disruption is the process by which a star is torn apart by the strong gravitational force of a nearby massive or supermassive black hole. About half of the star's debris remains bound to the black hole, and is ultimately accreted. It is predicted \cite{Hills75,Rees:1988bf,Lacy82,Phinney89} that  \td\ with the highest mass accretion rate 
should generate a relativistic jet. This jet can accelerate protons or nuclei to ultra-high energies \cite{Farrar:2008ex,Farrar:2014yla}, with \ns\ expected as a byproduct \cite{Wang:2011ip,Murase:2013ffa}. 
To date, three jet-hosting (``jetted'') \td\ have been robustly identified in X-rays observations \cite{Burrows:2011dn,Cenko:2011ys,Brown:2015amy} (see also \cite{vanVelzen:2015rlh,Lei:2015fbf}). Overall, they are consistent with the disruption of a main sequence star by a supermassive black hole (\sbh, $M > 10^5 \msun$, see \eg\ \cite{Burrows:2011dn,Bloom:2011xk}), but the disruption of a white dwarf (\wdw) star by an intermediate mass black hole (\bh, $M \sim (10^3 - 10^5) \msun$) is in principle a viable explanation \cite{Krolik:2011fb}. Regardless of the specific interpretation of observations, it is natural to expect diversity in the population of \td, involving black holes spanning many orders of magnitude in mass, as well as different types of stars.

\begin{figure*}[t!]
  \centering
  \includegraphics[width=0.4\textwidth]{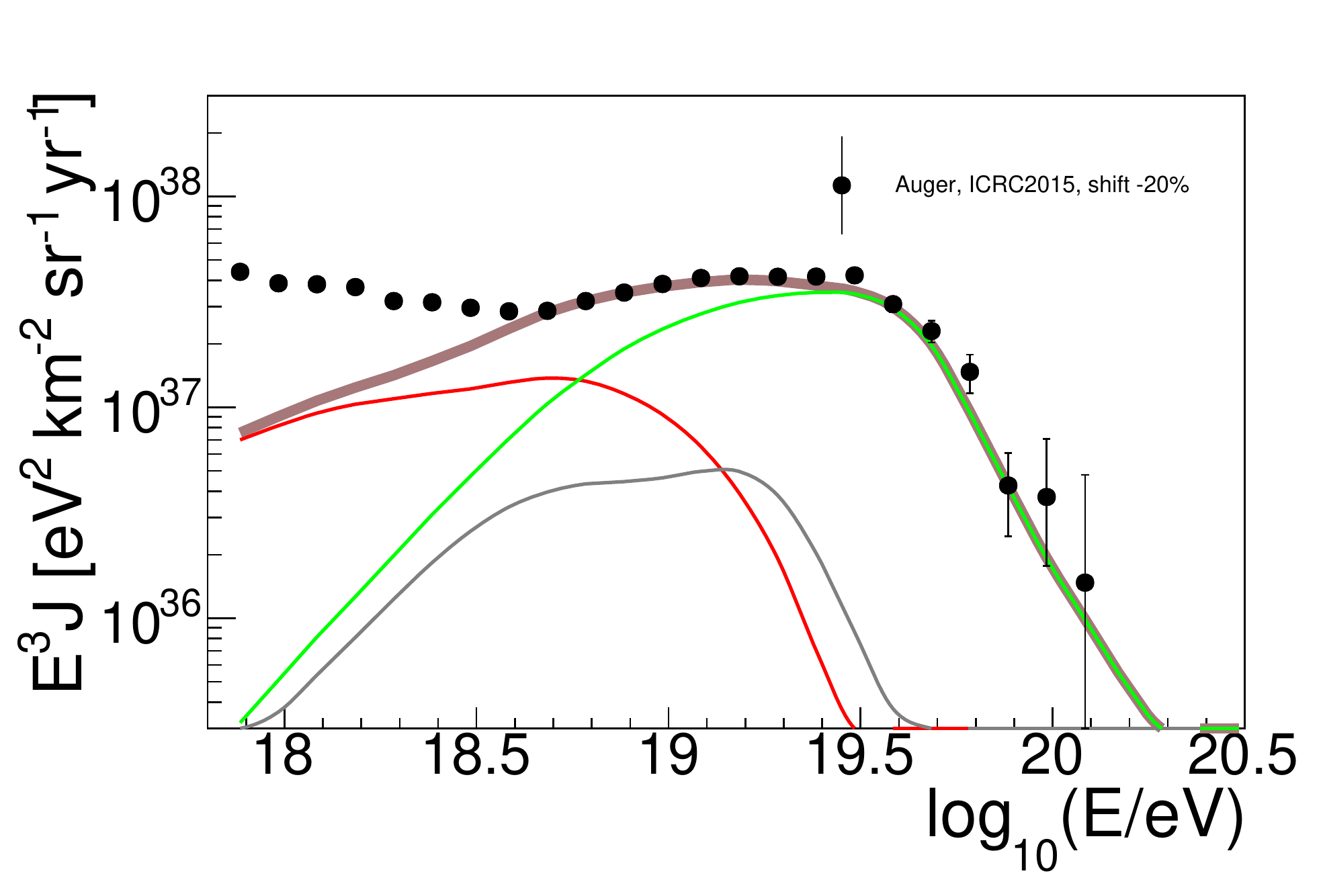}
  \includegraphics[width=0.4\textwidth]{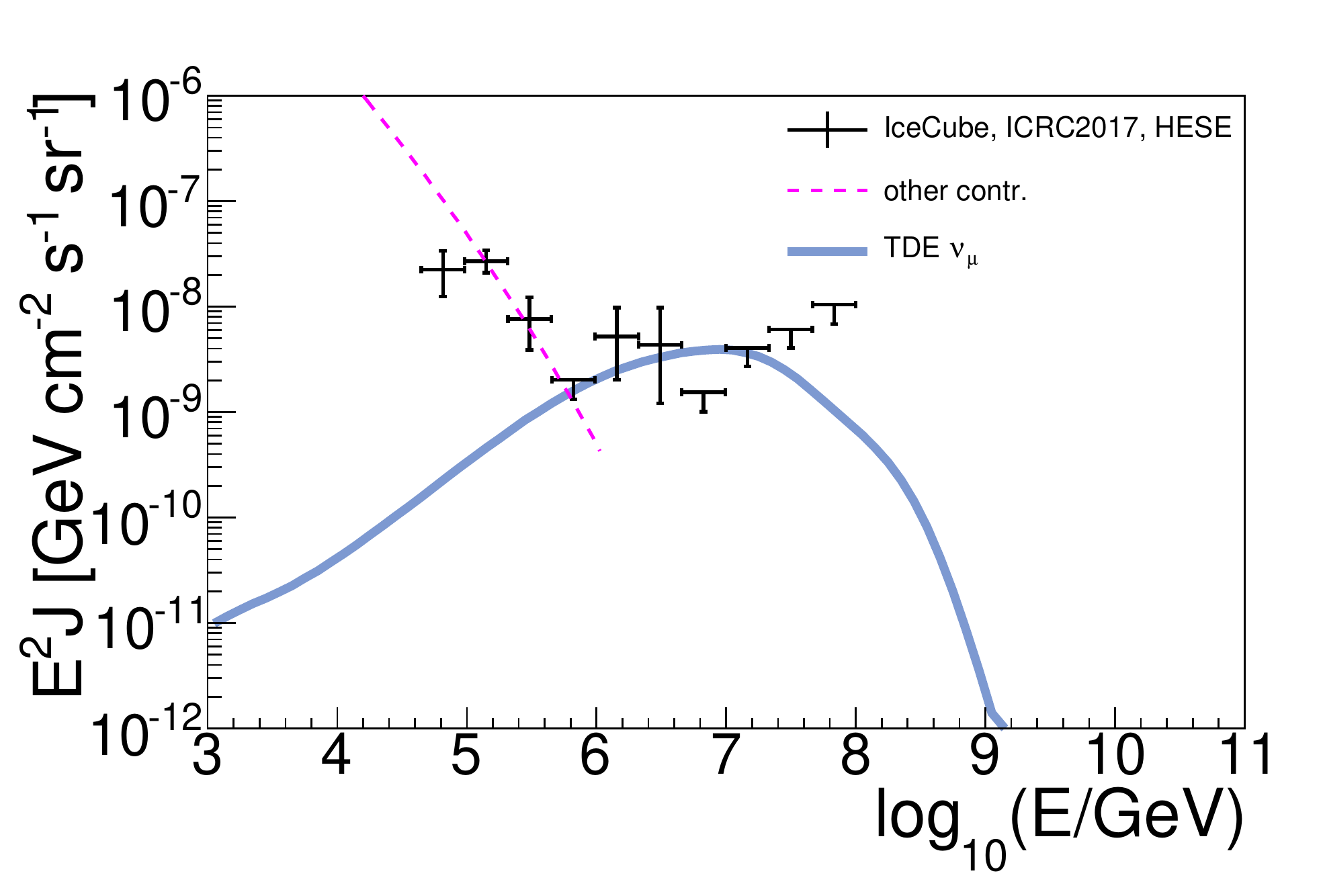}
  \\
\includegraphics[width=0.8\textwidth]{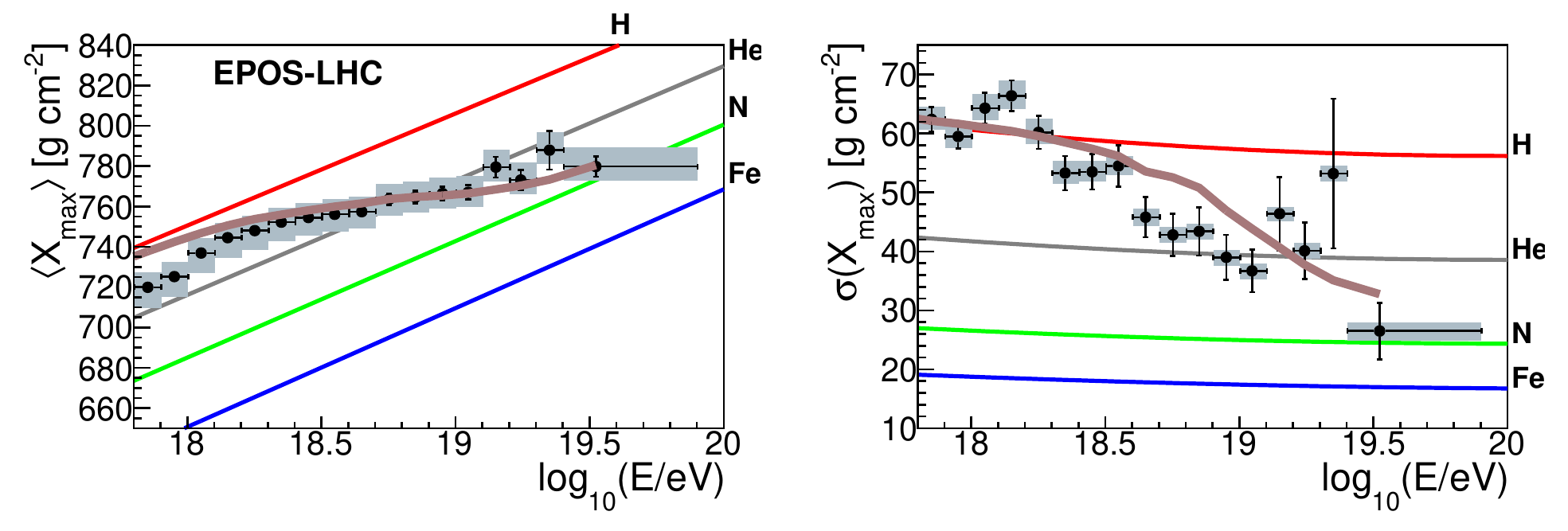}
\caption{
Cosmic ray and neutrino observables corresponding to a parameter space point describing both UHECR and neutrino data at the highest energies (point A in \figu{fit}, $L_X=10^{47}$ erg/s, $R=10^{9.6}$ km, with $G=540$). 
{\it Upper right panel}: predicted muon neutrino spectrum from \td,  compared to the data from the High Energy Starting Events at IceCube~\cite{kopper:2017}. An additional flux, which might be of atmospheric origin (taken from \cite{kopper:2017}), is also shown. {\it Upper left panel}: Simulated energy spectrum of UHECRs (thick curve); and its components from (groups of) different nuclear species (thin, same color coding as in the bottom panels). For comparison, the Auger data are shown~\cite{Valino:2015}.
{\it Lower panels}: Predictions and data~\cite{Porcelli:2015} on the average (left) and standard deviation (right) of the $X_{\mathrm{max}}$ distributions as a function of the energy. For predictions, EPOS-LHC~\cite{Pierog:2013ria} is assumed as the interaction model for UHECR-air interactions. A shift  of $-20\%$ is applied to the energy scale of all the \uh\ data, see text. 
 }\label{fig:bestfit}
\vspace{-0.3cm}
\end{figure*}

TDEs as the sources of extragalactic neutrinos~\cite{Wang:2011ip, Wang:2015mmh,Dai:2016gtz,Senno:2016bso,Lunardini:2016xwi} and UHECRs~\cite{AlvesBatista:2017shr,Zhang:2017hom} have been recently very actively discussed  in the literature. 
Notably,  \Refs~\cite{AlvesBatista:2017shr,Zhang:2017hom} focused on the recent observation of a mixed nuclear composition of UHECRs by the Pierre Auger Observatory~\cite{Aab:2014kda}. They  discussed how  \td\ offer an attractive and natural explanation of the composition if the disrupted stars have mid-to-heavy compositions; WDs were proposed as ideal candidates.   
So far, a consistent study of the joint production of UHECRs and \ns\ in the jet generated by tidal disruption has not been performed. In a detailed discussion, Zhang et al. \cite{Zhang:2017hom} concluded that the most prominently used scenario, the internal shock model, faces the difficulty that nuclei will disintegrate \emph{in the jet}, leading to a complex pattern of production of secondary nuclei and neutrinos. As a consequence,  it has been assumed that the UHECRs come from external shocks (or from regions with low enough radiation densities) for the sake of simplicity.

In this work, we present the first consistent calculation of neutrino and UHECR production in TDE jets in the internal shock scenario. Our main purpose is to demonstrate that TDEs, with appropriate nuclear injection composition,  are a viable common origin for the \ns\ and UHECRs.  The nuclear cascade in the source is modeled explicitly, using techniques that have been successfully applied before to GRBs~\cite{Boncioli:2016lkt,Biehl:2017zlw}. 


Here the methodology is outlined briefly; more details are given in the supplemental material. 
We model the TDE jet emission numerically,  following \Ref~\cite{Lunardini:2016xwi}  in the choice of the jet parameters, which are inspired by the best observed jetted TDE, Swift J1644+57~\cite{Burrows:2011dn}.   For the sake of simplicity, we assume that a single nuclear species, $\mathrm{{}^{14}N}$, is injected in the jet. This pure injection composition has been found to approximate the results obtained with a mixed carbon-oxygen (C-O) injection, which might be expected in the disruption of a C-O WD.  
This choice is also inspired by the recent observations of nitrogen emission lines in TDE observations \cite{Cenko:2016pum,Brown:2017sad}. 
Other possibilities for the nuclear composition, including   ONeMg dwarfs from past supernovae or WDs with explosive nuclear burning (see \eg\ \cite{Zhang:2017hom}), are other options which will not be considered here for brevity.
 
We simulate the interactions in the TDE jet with the {\it NeuCosmA} code as in \cite{Biehl:2017zlw}. The resulting cosmic ray and neutrino spectra are then processed by the {\it SimProp} code \cite{Aloisio:2017iyh}, which models the \uh\ propagation through the extragalactic space, and also computes the cosmogenic neutrino flux.  The mechanism for the escape of the cosmic rays from the sources is calculated as in \Ref~\cite{Baerwald:2013pu}, leading to hard spectra ejected from the source and injected  in the extragalactic space. These spectra are compatible with the results from the UHECR global fit by the Auger Collaboration~\cite{Aab:2016zth} (depending on the source evolution). 
We obtain the diffuse particle fluxes at Earth, using the assumption that all TDE jets are identical in the cosmologically co-moving frame, and that their rate evolves negatively with the redshift (approximately as $\sim (1+z)^{-3}$), following the evolution of the  number density of SMBHs as calculated in \Ref~\cite{Shankar:2007zg} (see also \cite{Stone:2014wxa,Kochanek:2016zzg,Lunardini:2016xwi}). We also compute the first two moments of the distributions of the quantity $X_{\mathrm{max}}$, which is defined as 
 the depth at which the energy deposited in the atmosphere by a cosmic ray shower reaches its maximum; $X_{\mathrm{max}}$  depends strongly on the mass of the primary cosmic ray nucleus.

\begin{figure*}[t!]
 \includegraphics[width=0.47\textwidth]{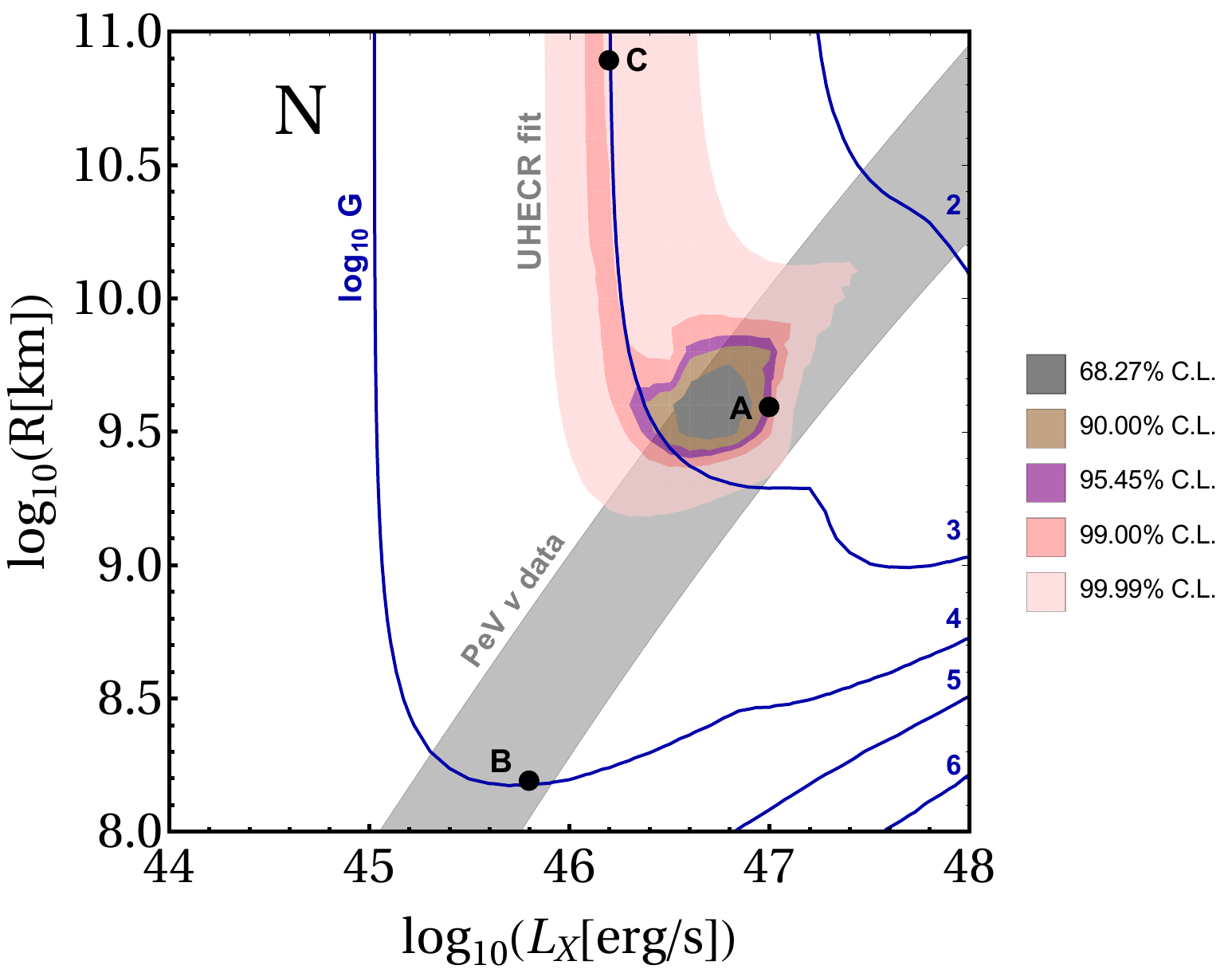} \hspace{0.4cm} 
\includegraphics[width=0.38\textwidth]{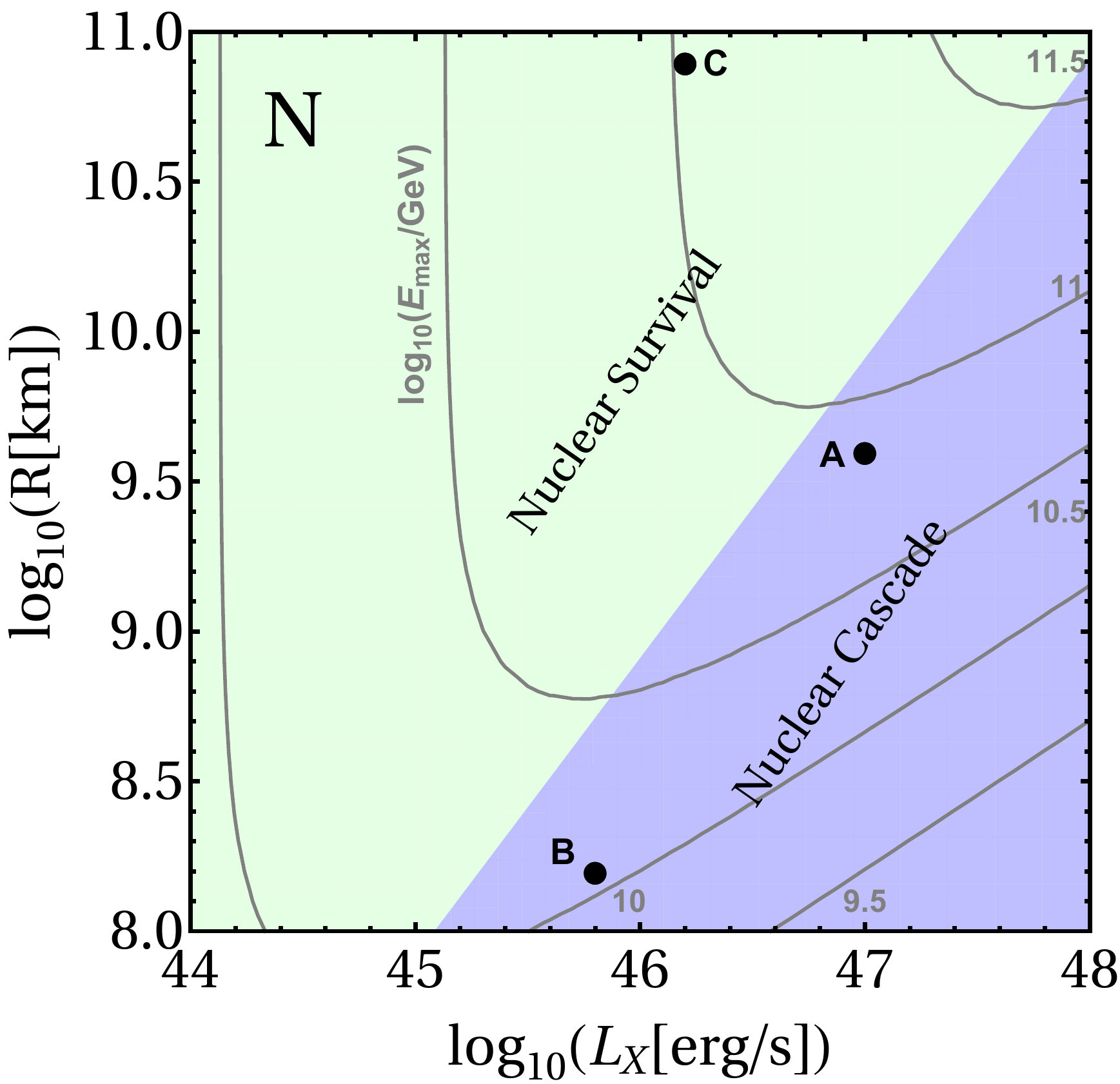}
\caption{{\it Left panel}: Results of the fit to UHECR and the description of PeV neutrino data
as a function of $L_X$ and $R$ (shaded contours, CL for two parameters).  The curves show isocontours of $\log_{10} G$  (see \equ{g}) obtained from the cosmic ray fit.  For each point $(L_X, R)$, the value of $G$ that maximizes the likelihood is used, \ie, $G$ is marginalized in the fit. 
{\it Right panel}: Different regimes in the parameter space for the nuclear cascade to develop in the source (shaded regions), as discussed in the main text. The curves show  $\log_{10}(E_{\mathrm{max}}/\mathrm{GeV})$, with $E_{\text{max}}$ being the obtained maximal energy in the observer's frame.  }\label{fig:fit}
\end{figure*}

To assess the compatibility with observations, we analyze the  Pierre Auger Observatory  data for the \uh\ spectrum \cite{Valino:2015} and for the distributions of $X_{\text{max}}$ \cite{Aab:2014kda} beyond $10^{19} \, \mathrm{eV}$. A fit of these data is performed, including a downshift (of the data) of 20\% in the energy scale to better match the maximal energy of the spectrum. The shift amount is comparable to the energy scale uncertainty of the Auger experiment (14\%). It is treated as experimental systematics here, but it is degenerate with the acceleration efficiency (or even nuclear injection composition) of the primaries, which can be adjusted accordingly to reach high enough maximal energies. After the \uh\ fit, as a separate step, we check the compatibility of the results with the IceCube neutrino data (measured data points beyond PeV energies~\cite{kopper:2017}).

The \uh\ fit is performed  using the maximum likelihood method, with three fit parameters: the production radius $R$ (distance from black hole where internal shocks occur), the X-ray luminosity $L_X$, and a single normalization parameter, $G$. 
 The latter takes into account the degeneracy between the  baryonic loading $\xi_A$ -- defined as the energy injected as nuclei over the total X-ray energy in the Swift range 0.4-13.5~keV~\cite{Burrows:2011dn} -- and the local apparent rate of jetted TDEs $\tilde R(0)$. It is defined as
\begin{equation}
  G \equiv \xi_A \times \frac{\tilde R(0)}{0.1 \, \mathrm{Gpc^{-3}} \, \mathrm{yr^{-1}}}~.
\label{equ:g}
\end{equation}
The reference value chosen for $\tilde R(0)$ is the rate of WD-IMBH disruptions inferred from observations~\cite{Senno:2016bso,Zhang:2017hom} $\tilde R(0) \sim 0.01- 0.1 \, \mathrm{Gpc^{-3}} \, \mathrm{yr^{-1}}$ (which is  in agreement with theoretical arguments, see \eg\ \cite{Krolik:2011fb}).

\figu{bestfit} shows our result for a parameter space point fitting UHECR (upper left panel and lower panels) and describing the PeV neutrino data (upper right panel). One can easily see that the UHECR spectrum and $X_{\mathrm{max}}$ beyond $10^{18.7} \, \mathrm{eV}$, and the neutrino spectrum at PeV energies are reproduced very well. The lower energy \n\ flux is underestimated in our model; the additional flux needed to reproduce the data in this region might be of atmospheric origin~\cite{kopper:2017}, of Galactic origin, or from a different source class (see \eg\   \Ref~\cite{Palladino:2016xsy}).  We emphasize that, for the parameters in  \figu{bestfit},  the source is optically thick to photo-hadronic interactions at the highest energies. Therefore, the effect of nuclear disintegration in the source  is important here. By including systematics (energy calibration error), we  obtain a better fit compared to  \Ref~\cite{Zhang:2017hom}, where the  UHECR data  are described in the nuclear survival regime for the disruption of C-O WDs, and a poor fit to the energy spectrum is found.

In \figu{fit}, left panel, we show (filled) the confidence level contours for the  fit to the \uh\ data, in the space of $L_X$ and $R$, after marginalizing over $G$; iso-contours of $\log_{10} G$ at the minimum are shown as well. We also superimpose the region where the predicted \n\ flux  is within $1\sigma$ from the two PeV data points of IceCube, thus providing an acceptable description of them. Point A in the figure gives the parameters used in \figu{bestfit}; point~B marks the best description of the PeV neutrino data, and point~C corresponds to a reasonable fit to the UHECRs in a different physics regime. For points B and C, only one data set (\uh\ or \ns) can be described well, but not both. Note that for the \uh\ data the statistical errors are smaller than the systematics ones; 
however, we find that the 99.99\% CL region in \figu{fit} is wide enough to be representative of the fit results that can be obtained if systematics (such as on the cosmic ray propagation model, as discussed for example in \cite{Batista:2015mea,Aab:2016zth}) are included (see also supplemental material).

In order to understand what physics determines the allowed regions found in the fit, we show the different regimes of the nuclear cascade in right panel of \figu{fit}: the one where the collision region is optically thick to nuclear disintegration (``nuclear cascade'')~\cite{Biehl:2017zlw}, and the complementary one where disintegration is inefficient (``nuclear survival'').  Iso-countours of the maximal energy of the nuclei spectrum, $E_{\text{max}}$, are shown as well. 
It appears that the allowed region of the \uh\ fit mainly follows the contour  $E_{\text{max}} \simeq 10^{10.8}~{\rm GeV}$ for the maximum energy in the source; this value  indeed reproduces the UHE range observed of cosmic rays at at Earth. 
Instead, the region preferred by the PeV neutrino data correlates with the nuclear cascade region, because nuclear disintegration and \n\ production require similar (but not too high) radiation densities for the photo-nuclear processes. At point C, the \uh\ spectrum and composition are reproduced but \n\ production is inefficient, thus resulting in a too low \n\ flux. At point B, the neutrino production is efficient enough to reach the level of the PeV data, but $E_{\text{max}}$ is too low, which means that the high energy \uh\ flux is not reproduced and the expected composition at Earth is heavier than what is measured. 

How physically plausible are the fit results? Let us discuss the best motivated scenario, which is the disruption of WDs.  We find that a good joint description is obtained for $G \simeq 540$ at point A, while  slightly lower values of the normalization parameter are obtained in the upper right part of the allowed region. 
In order to interpret the obtained value of $G$, assume that  50\% of the disrupted star's mass ($M_{\text{WD}} \sim M_{\odot}$) is efficiently converted into the baryonic jet. For the parameters at point A and $\Gamma \sim 10$, one then obtains $\xi_A  \sim 2 \Gamma^2 M_{\mathrm{WD}}/E_X \simeq 1800$ corresponding to a local apparent rate $\tilde R(0) \simeq 0.036 \, \mathrm{Gpc^{-3}} \, \mathrm{yr^{-1}}$ -- see \equ{g} -- which is loosely consistent with the rate of \wdw\ disruption. 
Higher local rates can be compensated by a lower baryonic loading, which means that a smaller fraction of energy of the disrupted star is converted into the jet. Note that additional lower limits on the apparent local rate may come from negative searches from neutrino multiplets from the same source \cite{Kowalski:2014zda,Ahlers:2014ioa,Murase:2016gly}.

As far as the limitations of our model are concerned, 
one uncertainty is the evolution of the TDE rate with redshift. A realistic application to WDs would require modeling the rate of \wdw\ disruption  by IMBHs, which depends on the unknown redshift evolution of the \bh\ number density. The negative evolution used here may apply; however, recent studies suggest that black holes of intermediate and small mass might be less numerous today than in the past, due to having merged into more massive black holes \cite{Alexander:2017arl}. This would suggest a less negative, or even positive evolution of the TDE rate \cite{alexander} with redshift. We have checked that the combined description of UHECR and PeV neutrino data rapidly becomes more challenging with the evolution becoming more positive (see supplemental material). 
   
A step towards a more realistic model would entail varying the input parameters that here have been kept fixed and inspired by the Swift J1644+57 observation. An alternative choice of parameters is offered by the idea that ultra-long GRBs might be caused by the disruption of WDs \cite{Shcherbakov:2012zt,Ioka:2016yzc} (see, however, \cite{Greiner:2015lia}), with GRB 111209A being a candidate.  Compared to Swift J1644+57, these bursts have a shorter duration and different X-ray spectra \cite{Levan:2013gcz}, and possibly a shorter variability time scale \cite{MacLeod:2014mha}. Another option is the tidal disruption of neutron stars, which may be associated to gravitational wave events such as GW170817~\cite{TheLIGOScientific:2017qsa}. For example, the observed short GRB in the follow-up of this event~\cite{Monitor:2017mdv}, SGRB 170817A, may be interpreted as a representative of a new population of jetted TDEs then.
More broadly, one may consider multiple classes of sources as contributing to the UHECR and \n\ fluxes. For example, a scenario including disruptions of both main sequence stars and WDs could lead to an even better description of the PeV \n\ spectrum, with main sequence stars reproducing the part of the spectrum at lower energies \cite{Lunardini:2016xwi}.


In summary, we have demonstrated that TDE jets with mid-to-heavy nuclear composition can reproduce both the observed cosmic rays and neutrinos at the highest energies, with typical parameters $L_X \simeq 10^{46}$ to $10^{47} \, \mathrm{erg \, s^{-1}}$, and $R \simeq 10^{9.5} \, \mathrm{km}$ (distance of production region from black hole).  We find that two important ingredients are necessary for a common description:  the first is that nuclear disintegration should be efficient \emph{in the jet}. This is because efficient neutrino production requires high radiation densities, which in turn implies efficient disintegration of nuclei. Therefore, the nuclear cascade in the jet has to be computed, and this computation is a key novelty of our work for TDEs.  The second condition is that the evolution of the sources with redshift should be negative (\ie, the jets should be less frequent in the past than today); indeed, this evolution is known to lead to a good fit of the \uh\ data~\cite{Taylor:2015rla}.  It is plausible for TDEs following the SMBH mass function, but debated for intermediate black hole masses. A consequence of the negative source evolution is that cosmogenic neutrinos will not be be detected, neither in the current nor in the next generation of experiments (see supplementary materials).
   The two conditions greatly restrict the classes of objects that can host such jets: the first requires large masses ($M \sim 0.1 - 1 \msun$) of material with heavy composition, and the second excludes many sources (such as supernovae or the GRBs) which are expected to track the star formation rate.  The disruption of WDs by massive black holes appears as a natural realization because of  their carbon-oxygen composition.  We have also demonstrated that in our description the  local apparent source density and the baryonic loading of the jets are degenerate,  which leaves room for various interpretations. 
   
We conclude that future observations will help to substantiate, or disfavor, the TDE origin of UHECRs and \ns. In particular, a higher number of precision observations of jetted \td\ will help to constrain the jet parameters.  An association of neutrinos or UHECRs with \td\ could be obtained from multi-messenger studies, with cross correlation of  observations in time and -- when possible -- position in the sky. Finally, the recently observed short gamma-ray burst SGRB 170817A associated with the gravitational wave event GW 170817 may be indicative of a new class of ``tidal disruption events'' (if interpreted as black hole-neutron star merger) which may be interpreted in a similar framework.

\subsection*{Acknowledgments}

We thank  K. Murase and F. Oikonomou for useful discussions. 
CL is grateful to the DESY Zeuthen laboratory for hospitality when this project was initiated. She acknowledges funding from the National Science Foundation grant number PHY-1613708.
This work has been supported by the European Research Council (ERC) under the European Union’s Horizon 2020 research and innovation programme (Grant No. 646623).

\bibliographystyle{apsrev4-1}
\bibliography{references}

\begin{thebibliography}{64}
\expandafter\ifx\csname natexlab\endcsname\relax\def\natexlab#1{#1}\fi
\expandafter\ifx\csname bibnamefont\endcsname\relax
  \def\bibnamefont#1{#1}\fi
\expandafter\ifx\csname bibfnamefont\endcsname\relax
  \def\bibfnamefont#1{#1}\fi
\expandafter\ifx\csname citenamefont\endcsname\relax
  \def\citenamefont#1{#1}\fi
\expandafter\ifx\csname url\endcsname\relax
  \def\url#1{\texttt{#1}}\fi
\expandafter\ifx\csname urlprefix\endcsname\relax\def\urlprefix{URL }\fi
\providecommand{\bibinfo}[2]{#2}
\providecommand{\eprint}[2][]{\url{#2}}

\bibitem[{\citenamefont{Aartsen et~al.}(2013)}]{Aartsen:2013jdh}
\bibinfo{author}{\bibfnamefont{M.}~\bibnamefont{Aartsen}} \bibnamefont{et~al.}
  (\bibinfo{collaboration}{IceCube}), \bibinfo{journal}{Science}
  \textbf{\bibinfo{volume}{342}}, \bibinfo{pages}{1242856}
  (\bibinfo{year}{2013}), \eprint{1311.5238}.

\bibitem[{\citenamefont{Waxman and Bahcall}(1999)}]{Waxman:1998yy}
\bibinfo{author}{\bibfnamefont{E.}~\bibnamefont{Waxman}} \bibnamefont{and}
  \bibinfo{author}{\bibfnamefont{J.~N.} \bibnamefont{Bahcall}},
  \bibinfo{journal}{Phys. Rev.} \textbf{\bibinfo{volume}{D59}},
  \bibinfo{pages}{023002} (\bibinfo{year}{1999}), \eprint{hep-ph/9807282}.

\bibitem[{\citenamefont{Aartsen et~al.}(2017{\natexlab{a}})}]{Aartsen:2017wea}
\bibinfo{author}{\bibfnamefont{M.~G.} \bibnamefont{Aartsen}}
  \bibnamefont{et~al.} (\bibinfo{collaboration}{IceCube}),
  \bibinfo{journal}{Astrophys. J.} \textbf{\bibinfo{volume}{843}},
  \bibinfo{pages}{112} (\bibinfo{year}{2017}{\natexlab{a}}),
  \eprint{1702.06868}.

\bibitem[{\citenamefont{Aartsen et~al.}(2017{\natexlab{b}})}]{Aartsen:2016lir}
\bibinfo{author}{\bibfnamefont{M.~G.} \bibnamefont{Aartsen}}
  \bibnamefont{et~al.} (\bibinfo{collaboration}{IceCube}),
  \bibinfo{journal}{Astrophys. J.} \textbf{\bibinfo{volume}{835}},
  \bibinfo{pages}{45} (\bibinfo{year}{2017}{\natexlab{b}}),
  \eprint{1611.03874}.

\bibitem[{\citenamefont{Murase et~al.}(2013)\citenamefont{Murase, Ahlers, and
  Lacki}}]{Murase:2013rfa}
\bibinfo{author}{\bibfnamefont{K.}~\bibnamefont{Murase}},
  \bibinfo{author}{\bibfnamefont{M.}~\bibnamefont{Ahlers}}, \bibnamefont{and}
  \bibinfo{author}{\bibfnamefont{B.~C.} \bibnamefont{Lacki}},
  \bibinfo{journal}{Phys.Rev.} \textbf{\bibinfo{volume}{D88}},
  \bibinfo{pages}{121301} (\bibinfo{year}{2013}), \eprint{1306.3417}.

\bibitem[{\citenamefont{Bechtol et~al.}(2017)\citenamefont{Bechtol, Ahlers,
  Di~Mauro, Ajello, and Vandenbroucke}}]{Bechtol:2015uqb}
\bibinfo{author}{\bibfnamefont{K.}~\bibnamefont{Bechtol}},
  \bibinfo{author}{\bibfnamefont{M.}~\bibnamefont{Ahlers}},
  \bibinfo{author}{\bibfnamefont{M.}~\bibnamefont{Di~Mauro}},
  \bibinfo{author}{\bibfnamefont{M.}~\bibnamefont{Ajello}}, \bibnamefont{and}
  \bibinfo{author}{\bibfnamefont{J.}~\bibnamefont{Vandenbroucke}},
  \bibinfo{journal}{Astrophys. J.} \textbf{\bibinfo{volume}{836}},
  \bibinfo{pages}{47} (\bibinfo{year}{2017}), \eprint{1511.00688}.

\bibitem[{\citenamefont{Hills}(1975)}]{Hills75}
\bibinfo{author}{\bibfnamefont{J.~G.} \bibnamefont{Hills}},
  \bibinfo{journal}{Nature} \textbf{\bibinfo{volume}{254}},
  \bibinfo{pages}{295} (\bibinfo{year}{1975}),
  \urlprefix\url{http://dx.doi.org/10.1038/254295a0}.

\bibitem[{\citenamefont{Rees}(1988)}]{Rees:1988bf}
\bibinfo{author}{\bibfnamefont{M.~J.} \bibnamefont{Rees}},
  \bibinfo{journal}{Nature} \textbf{\bibinfo{volume}{333}},
  \bibinfo{pages}{523} (\bibinfo{year}{1988}).

\bibitem[{\citenamefont{{Lacy} et~al.}(1982)\citenamefont{{Lacy}, {Townes}, and
  {Hollenbach}}}]{Lacy82}
\bibinfo{author}{\bibfnamefont{J.~H.} \bibnamefont{{Lacy}}},
  \bibinfo{author}{\bibfnamefont{C.~H.} \bibnamefont{{Townes}}},
  \bibnamefont{and} \bibinfo{author}{\bibfnamefont{D.~J.}
  \bibnamefont{{Hollenbach}}}, \bibinfo{journal}{Astrophys. J.}
  \textbf{\bibinfo{volume}{262}}, \bibinfo{pages}{120} (\bibinfo{year}{1982}).

\bibitem[{\citenamefont{{Phinney}}(1989)}]{Phinney89}
\bibinfo{author}{\bibfnamefont{E.~S.} \bibnamefont{{Phinney}}}, in
  \emph{\bibinfo{booktitle}{The Center of the Galaxy}}, edited by
  \bibinfo{editor}{\bibfnamefont{M.}~\bibnamefont{{Morris}}}
  (\bibinfo{year}{1989}), vol. \bibinfo{volume}{136} of
  \emph{\bibinfo{series}{IAU Symposium}}, p. \bibinfo{pages}{543}.

\bibitem[{\citenamefont{Farrar and Gruzinov}(2009)}]{Farrar:2008ex}
\bibinfo{author}{\bibfnamefont{G.~R.} \bibnamefont{Farrar}} \bibnamefont{and}
  \bibinfo{author}{\bibfnamefont{A.}~\bibnamefont{Gruzinov}},
  \bibinfo{journal}{Astrophys. J.} \textbf{\bibinfo{volume}{693}},
  \bibinfo{pages}{329} (\bibinfo{year}{2009}), \eprint{0802.1074}.

\bibitem[{\citenamefont{Farrar and Piran}(2014)}]{Farrar:2014yla}
\bibinfo{author}{\bibfnamefont{G.~R.} \bibnamefont{Farrar}} \bibnamefont{and}
  \bibinfo{author}{\bibfnamefont{T.}~\bibnamefont{Piran}}
  (\bibinfo{year}{2014}), \eprint{1411.0704}.

\bibitem[{\citenamefont{Wang et~al.}(2011)\citenamefont{Wang, Liu, Dai, and
  Cheng}}]{Wang:2011ip}
\bibinfo{author}{\bibfnamefont{X.-Y.} \bibnamefont{Wang}},
  \bibinfo{author}{\bibfnamefont{R.-Y.} \bibnamefont{Liu}},
  \bibinfo{author}{\bibfnamefont{Z.-G.} \bibnamefont{Dai}}, \bibnamefont{and}
  \bibinfo{author}{\bibfnamefont{K.~S.} \bibnamefont{Cheng}},
  \bibinfo{journal}{Phys. Rev.} \textbf{\bibinfo{volume}{D84}},
  \bibinfo{pages}{081301} (\bibinfo{year}{2011}), \eprint{1106.2426}.

\bibitem[{\citenamefont{Murase and Ioka}(2013)}]{Murase:2013ffa}
\bibinfo{author}{\bibfnamefont{K.}~\bibnamefont{Murase}} \bibnamefont{and}
  \bibinfo{author}{\bibfnamefont{K.}~\bibnamefont{Ioka}}
  (\bibinfo{year}{2013}), \eprint{1306.2274}.

\bibitem[{\citenamefont{Burrows et~al.}(2011)}]{Burrows:2011dn}
\bibinfo{author}{\bibfnamefont{D.~N.} \bibnamefont{Burrows}}
  \bibnamefont{et~al.}, \bibinfo{journal}{Nature}
  \textbf{\bibinfo{volume}{476}}, \bibinfo{pages}{421} (\bibinfo{year}{2011}),
  \eprint{1104.4787}.

\bibitem[{\citenamefont{Cenko et~al.}(2012)}]{Cenko:2011ys}
\bibinfo{author}{\bibfnamefont{S.~B.} \bibnamefont{Cenko}}
  \bibnamefont{et~al.}, \bibinfo{journal}{Astrophys. J.}
  \textbf{\bibinfo{volume}{753}}, \bibinfo{pages}{77} (\bibinfo{year}{2012}),
  \eprint{1107.5307}.

\bibitem[{\citenamefont{Brown et~al.}(2015)\citenamefont{Brown, Levan, Stanway,
  Tanvir, Cenko, Berger, Chornock, and Cucchiaria}}]{Brown:2015amy}
\bibinfo{author}{\bibfnamefont{G.~C.} \bibnamefont{Brown}},
  \bibinfo{author}{\bibfnamefont{A.~J.} \bibnamefont{Levan}},
  \bibinfo{author}{\bibfnamefont{E.~R.} \bibnamefont{Stanway}},
  \bibinfo{author}{\bibfnamefont{N.~R.} \bibnamefont{Tanvir}},
  \bibinfo{author}{\bibfnamefont{S.~B.} \bibnamefont{Cenko}},
  \bibinfo{author}{\bibfnamefont{E.}~\bibnamefont{Berger}},
  \bibinfo{author}{\bibfnamefont{R.}~\bibnamefont{Chornock}}, \bibnamefont{and}
  \bibinfo{author}{\bibfnamefont{A.}~\bibnamefont{Cucchiaria}},
  \bibinfo{journal}{Mon. Not. Roy. Astron. Soc.}
  \textbf{\bibinfo{volume}{452}}, \bibinfo{pages}{4297} (\bibinfo{year}{2015}),
  \eprint{1507.03582}.

\bibitem[{\citenamefont{van Velzen et~al.}(2016)}]{vanVelzen:2015rlh}
\bibinfo{author}{\bibfnamefont{S.}~\bibnamefont{van Velzen}}
  \bibnamefont{et~al.}, \bibinfo{journal}{Science}
  \textbf{\bibinfo{volume}{351}}, \bibinfo{pages}{62} (\bibinfo{year}{2016}),
  \eprint{1511.08803}.

\bibitem[{\citenamefont{Lei et~al.}(2016)\citenamefont{Lei, Yuan, Zhang, and
  Wang}}]{Lei:2015fbf}
\bibinfo{author}{\bibfnamefont{W.-H.} \bibnamefont{Lei}},
  \bibinfo{author}{\bibfnamefont{Q.}~\bibnamefont{Yuan}},
  \bibinfo{author}{\bibfnamefont{B.}~\bibnamefont{Zhang}}, \bibnamefont{and}
  \bibinfo{author}{\bibfnamefont{D.}~\bibnamefont{Wang}},
  \bibinfo{journal}{Astrophys. J.} \textbf{\bibinfo{volume}{816}},
  \bibinfo{pages}{20} (\bibinfo{year}{2016}), \eprint{1511.01206}.

\bibitem[{\citenamefont{Bloom et~al.}(2011)}]{Bloom:2011xk}
\bibinfo{author}{\bibfnamefont{J.~S.} \bibnamefont{Bloom}}
  \bibnamefont{et~al.}, \bibinfo{journal}{Science}
  \textbf{\bibinfo{volume}{333}}, \bibinfo{pages}{203} (\bibinfo{year}{2011}),
  \eprint{1104.3257}.

\bibitem[{\citenamefont{Krolik and Piran}(2011)}]{Krolik:2011fb}
\bibinfo{author}{\bibfnamefont{J.~H.} \bibnamefont{Krolik}} \bibnamefont{and}
  \bibinfo{author}{\bibfnamefont{T.}~\bibnamefont{Piran}},
  \bibinfo{journal}{Astrophys. J.} \textbf{\bibinfo{volume}{743}},
  \bibinfo{pages}{134} (\bibinfo{year}{2011}), \eprint{1106.0923}.

\bibitem[{\citenamefont{{Kopper, C. et al.}}(2017)}]{kopper:2017}
\bibinfo{author}{\bibnamefont{{Kopper, C. et al.}}}
  (\bibinfo{collaboration}{IceCube Collaboration})
  (\bibinfo{publisher}{PoS(ICRC2017)981}, \bibinfo{year}{2017}).

\bibitem[{\citenamefont{{Vali\~no, I. et al.}}(2015)}]{Valino:2015}
\bibinfo{author}{\bibnamefont{{Vali\~no, I. et al.}}}
  (\bibinfo{collaboration}{Pierre Auger Collaboration})
  (\bibinfo{publisher}{PoS(ICRC2015)271}, \bibinfo{year}{2015}).

\bibitem[{\citenamefont{{Porcelli, A. et al.}}(2015)}]{Porcelli:2015}
\bibinfo{author}{\bibnamefont{{Porcelli, A. et al.}}}
  (\bibinfo{collaboration}{Pierre Auger Collaboration})
  (\bibinfo{publisher}{PoS(ICRC2015)420}, \bibinfo{year}{2015}).

\bibitem[{\citenamefont{Pierog et~al.}(2015)\citenamefont{Pierog, Karpenko,
  Katzy, Yatsenko, and Werner}}]{Pierog:2013ria}
\bibinfo{author}{\bibfnamefont{T.}~\bibnamefont{Pierog}},
  \bibinfo{author}{\bibfnamefont{I.}~\bibnamefont{Karpenko}},
  \bibinfo{author}{\bibfnamefont{J.~M.} \bibnamefont{Katzy}},
  \bibinfo{author}{\bibfnamefont{E.}~\bibnamefont{Yatsenko}}, \bibnamefont{and}
  \bibinfo{author}{\bibfnamefont{K.}~\bibnamefont{Werner}},
  \bibinfo{journal}{Phys. Rev.} \textbf{\bibinfo{volume}{C92}},
  \bibinfo{pages}{034906} (\bibinfo{year}{2015}), \eprint{1306.0121}.

\bibitem[{\citenamefont{Wang and Liu}(2016)}]{Wang:2015mmh}
\bibinfo{author}{\bibfnamefont{X.-Y.} \bibnamefont{Wang}} \bibnamefont{and}
  \bibinfo{author}{\bibfnamefont{R.-Y.} \bibnamefont{Liu}},
  \bibinfo{journal}{Phys. Rev.} \textbf{\bibinfo{volume}{D93}},
  \bibinfo{pages}{083005} (\bibinfo{year}{2016}), \eprint{1512.08596}.

\bibitem[{\citenamefont{Dai and Fang}(2017)}]{Dai:2016gtz}
\bibinfo{author}{\bibfnamefont{L.}~\bibnamefont{Dai}} \bibnamefont{and}
  \bibinfo{author}{\bibfnamefont{K.}~\bibnamefont{Fang}},
  \bibinfo{journal}{Mon. Not. Roy. Astron. Soc.}
  \textbf{\bibinfo{volume}{469}}, \bibinfo{pages}{1354} (\bibinfo{year}{2017}),
  \eprint{1612.00011}.

\bibitem[{\citenamefont{Senno et~al.}(2017)\citenamefont{Senno, Murase, and
  Meszaros}}]{Senno:2016bso}
\bibinfo{author}{\bibfnamefont{N.}~\bibnamefont{Senno}},
  \bibinfo{author}{\bibfnamefont{K.}~\bibnamefont{Murase}}, \bibnamefont{and}
  \bibinfo{author}{\bibfnamefont{P.}~\bibnamefont{Meszaros}},
  \bibinfo{journal}{Astrophys. J.} \textbf{\bibinfo{volume}{838}},
  \bibinfo{pages}{3} (\bibinfo{year}{2017}), \eprint{1612.00918}.

\bibitem[{\citenamefont{Lunardini and Winter}(2017)}]{Lunardini:2016xwi}
\bibinfo{author}{\bibfnamefont{C.}~\bibnamefont{Lunardini}} \bibnamefont{and}
  \bibinfo{author}{\bibfnamefont{W.}~\bibnamefont{Winter}},
  \bibinfo{journal}{Phys. Rev.} \textbf{\bibinfo{volume}{D95}},
  \bibinfo{pages}{123001} (\bibinfo{year}{2017}), \eprint{1612.03160}.

\bibitem[{\citenamefont{Alves~Batista and Silk}(2017)}]{AlvesBatista:2017shr}
\bibinfo{author}{\bibfnamefont{R.}~\bibnamefont{Alves~Batista}}
  \bibnamefont{and} \bibinfo{author}{\bibfnamefont{J.}~\bibnamefont{Silk}}
  (\bibinfo{year}{2017}), \eprint{1702.06978}.

\bibitem[{\citenamefont{Zhang et~al.}(2017)\citenamefont{Zhang, Murase,
  Oikonomou, and Li}}]{Zhang:2017hom}
\bibinfo{author}{\bibfnamefont{B.~T.} \bibnamefont{Zhang}},
  \bibinfo{author}{\bibfnamefont{K.}~\bibnamefont{Murase}},
  \bibinfo{author}{\bibfnamefont{F.}~\bibnamefont{Oikonomou}},
  \bibnamefont{and} \bibinfo{author}{\bibfnamefont{Z.}~\bibnamefont{Li}}
  (\bibinfo{year}{2017}), \eprint{1706.00391}.

\bibitem[{\citenamefont{Aab et~al.}(2014)}]{Aab:2014kda}
\bibinfo{author}{\bibfnamefont{A.}~\bibnamefont{Aab}} \bibnamefont{et~al.}
  (\bibinfo{collaboration}{Pierre Auger Collaboration}),
  \bibinfo{journal}{Phys. Rev.} \textbf{\bibinfo{volume}{D90}},
  \bibinfo{pages}{122005} (\bibinfo{year}{2014}), \eprint{1409.4809}.

\bibitem[{\citenamefont{Boncioli et~al.}(2017)\citenamefont{Boncioli,
  Fedynitch, and Winter}}]{Boncioli:2016lkt}
\bibinfo{author}{\bibfnamefont{D.}~\bibnamefont{Boncioli}},
  \bibinfo{author}{\bibfnamefont{A.}~\bibnamefont{Fedynitch}},
  \bibnamefont{and} \bibinfo{author}{\bibfnamefont{W.}~\bibnamefont{Winter}},
  \bibinfo{journal}{Scientific Reports} \textbf{\bibinfo{volume}{7}},
  \bibinfo{pages}{4882} (\bibinfo{year}{2017}), \eprint{1607.07989}.

\bibitem[{\citenamefont{Biehl et~al.}(accepted)\citenamefont{Biehl, Boncioli,
  Fedynitch, and Winter}}]{Biehl:2017zlw}
\bibinfo{author}{\bibfnamefont{D.}~\bibnamefont{Biehl}},
  \bibinfo{author}{\bibfnamefont{D.}~\bibnamefont{Boncioli}},
  \bibinfo{author}{\bibfnamefont{A.}~\bibnamefont{Fedynitch}},
  \bibnamefont{and} \bibinfo{author}{\bibfnamefont{W.}~\bibnamefont{Winter}},
  \bibinfo{journal}{A\&A}  (\bibinfo{year}{accepted}), \eprint{1705.08909}.

\bibitem[{\citenamefont{Cenko et~al.}(2016)}]{Cenko:2016pum}
\bibinfo{author}{\bibfnamefont{S.~B.} \bibnamefont{Cenko}}
  \bibnamefont{et~al.}, \bibinfo{journal}{Astrophys. J.}
  \textbf{\bibinfo{volume}{818}}, \bibinfo{pages}{L32} (\bibinfo{year}{2016}),
  \eprint{1601.03331}.

\bibitem[{\citenamefont{Brown et~al.}(2017)}]{Brown:2017sad}
\bibinfo{author}{\bibfnamefont{J.~S.} \bibnamefont{Brown}} \bibnamefont{et~al.}
  (\bibinfo{year}{2017}), \eprint{1704.02321}.

\bibitem[{\citenamefont{Aloisio et~al.}(2017)\citenamefont{Aloisio, Boncioli,
  Di~Matteo, Grillo, Petrera, and Salamida}}]{Aloisio:2017iyh}
\bibinfo{author}{\bibfnamefont{R.}~\bibnamefont{Aloisio}},
  \bibinfo{author}{\bibfnamefont{D.}~\bibnamefont{Boncioli}},
  \bibinfo{author}{\bibfnamefont{A.}~\bibnamefont{Di~Matteo}},
  \bibinfo{author}{\bibfnamefont{A.~F.} \bibnamefont{Grillo}},
  \bibinfo{author}{\bibfnamefont{S.}~\bibnamefont{Petrera}}, \bibnamefont{and}
  \bibinfo{author}{\bibfnamefont{F.}~\bibnamefont{Salamida}}
  (\bibinfo{year}{2017}), \eprint{1705.03729, accepted by JCAP}.

\bibitem[{\citenamefont{Baerwald et~al.}(2013)\citenamefont{Baerwald,
  Bustamante, and Winter}}]{Baerwald:2013pu}
\bibinfo{author}{\bibfnamefont{P.}~\bibnamefont{Baerwald}},
  \bibinfo{author}{\bibfnamefont{M.}~\bibnamefont{Bustamante}},
  \bibnamefont{and} \bibinfo{author}{\bibfnamefont{W.}~\bibnamefont{Winter}},
  \bibinfo{journal}{Astrophys. J.} \textbf{\bibinfo{volume}{768}},
  \bibinfo{pages}{186} (\bibinfo{year}{2013}), \eprint{1301.6163}.

\bibitem[{\citenamefont{Aab et~al.}(2017)}]{Aab:2016zth}
\bibinfo{author}{\bibfnamefont{A.}~\bibnamefont{Aab}} \bibnamefont{et~al.}
  (\bibinfo{collaboration}{Pierre Auger}), \bibinfo{journal}{JCAP}
  \textbf{\bibinfo{volume}{1704}}, \bibinfo{pages}{038} (\bibinfo{year}{2017}),
  \eprint{1612.07155}.

\bibitem[{\citenamefont{Shankar et~al.}(2009)\citenamefont{Shankar, Weinberg,
  and Miralda-Escude}}]{Shankar:2007zg}
\bibinfo{author}{\bibfnamefont{F.}~\bibnamefont{Shankar}},
  \bibinfo{author}{\bibfnamefont{D.~H.} \bibnamefont{Weinberg}},
  \bibnamefont{and}
  \bibinfo{author}{\bibfnamefont{J.}~\bibnamefont{Miralda-Escude}},
  \bibinfo{journal}{Astrophys. J.} \textbf{\bibinfo{volume}{690}},
  \bibinfo{pages}{20} (\bibinfo{year}{2009}), \eprint{0710.4488}.

\bibitem[{\citenamefont{Stone and Metzger}(2016)}]{Stone:2014wxa}
\bibinfo{author}{\bibfnamefont{N.~C.} \bibnamefont{Stone}} \bibnamefont{and}
  \bibinfo{author}{\bibfnamefont{B.~D.} \bibnamefont{Metzger}},
  \bibinfo{journal}{Mon. Not. Roy. Astron. Soc.}
  \textbf{\bibinfo{volume}{455}}, \bibinfo{pages}{859} (\bibinfo{year}{2016}),
  \eprint{1410.7772}.

\bibitem[{\citenamefont{Kochanek}(2016)}]{Kochanek:2016zzg}
\bibinfo{author}{\bibfnamefont{C.~S.} \bibnamefont{Kochanek}},
  \bibinfo{journal}{Mon. Not. Roy. Astron. Soc.}
  \textbf{\bibinfo{volume}{461}}, \bibinfo{pages}{371} (\bibinfo{year}{2016}),
  \eprint{1601.06787}.

\bibitem[{\citenamefont{Palladino et~al.}(2016)\citenamefont{Palladino, Spurio,
  and Vissani}}]{Palladino:2016xsy}
\bibinfo{author}{\bibfnamefont{A.}~\bibnamefont{Palladino}},
  \bibinfo{author}{\bibfnamefont{M.}~\bibnamefont{Spurio}}, \bibnamefont{and}
  \bibinfo{author}{\bibfnamefont{F.}~\bibnamefont{Vissani}},
  \bibinfo{journal}{JCAP} \textbf{\bibinfo{volume}{1612}}, \bibinfo{pages}{045}
  (\bibinfo{year}{2016}), \eprint{1610.07015}.

\bibitem[{\citenamefont{Alves~Batista et~al.}(2015)\citenamefont{Alves~Batista,
  Boncioli, di~Matteo, van Vliet, and Walz}}]{Batista:2015mea}
\bibinfo{author}{\bibfnamefont{R.}~\bibnamefont{Alves~Batista}},
  \bibinfo{author}{\bibfnamefont{D.}~\bibnamefont{Boncioli}},
  \bibinfo{author}{\bibfnamefont{A.}~\bibnamefont{di~Matteo}},
  \bibinfo{author}{\bibfnamefont{A.}~\bibnamefont{van Vliet}},
  \bibnamefont{and} \bibinfo{author}{\bibfnamefont{D.}~\bibnamefont{Walz}},
  \bibinfo{journal}{JCAP} \textbf{\bibinfo{volume}{1510}}, \bibinfo{pages}{063}
  (\bibinfo{year}{2015}), \eprint{1508.01824}.

\bibitem[{\citenamefont{Kowalski}(2015)}]{Kowalski:2014zda}
\bibinfo{author}{\bibfnamefont{M.}~\bibnamefont{Kowalski}},
  \bibinfo{journal}{J. Phys. Conf. Ser.} \textbf{\bibinfo{volume}{632}},
  \bibinfo{pages}{012039} (\bibinfo{year}{2015}), \eprint{1411.4385}.

\bibitem[{\citenamefont{Ahlers and Halzen}(2014)}]{Ahlers:2014ioa}
\bibinfo{author}{\bibfnamefont{M.}~\bibnamefont{Ahlers}} \bibnamefont{and}
  \bibinfo{author}{\bibfnamefont{F.}~\bibnamefont{Halzen}},
  \bibinfo{journal}{Phys. Rev.} \textbf{\bibinfo{volume}{D90}},
  \bibinfo{pages}{043005} (\bibinfo{year}{2014}), \eprint{1406.2160}.

\bibitem[{\citenamefont{Murase and Waxman}(2016)}]{Murase:2016gly}
\bibinfo{author}{\bibfnamefont{K.}~\bibnamefont{Murase}} \bibnamefont{and}
  \bibinfo{author}{\bibfnamefont{E.}~\bibnamefont{Waxman}},
  \bibinfo{journal}{Phys. Rev.} \textbf{\bibinfo{volume}{D94}},
  \bibinfo{pages}{103006} (\bibinfo{year}{2016}), \eprint{1607.01601}.

\bibitem[{\citenamefont{{Alexander} and {Bar-Or}}(2017)}]{Alexander:2017arl}
\bibinfo{author}{\bibfnamefont{T.}~\bibnamefont{{Alexander}}} \bibnamefont{and}
  \bibinfo{author}{\bibfnamefont{B.}~\bibnamefont{{Bar-Or}}},
  \bibinfo{journal}{Nature Astronomy} \textbf{\bibinfo{volume}{1}},
  \bibinfo{eid}{0147} (\bibinfo{year}{2017}), \eprint{1701.00415}.

\bibitem[{\citenamefont{Alexander}()}]{alexander}
\bibinfo{author}{\bibfnamefont{T.}~\bibnamefont{Alexander}},
  \bibinfo{howpublished}{private communication}.

\bibitem[{\citenamefont{Shcherbakov et~al.}(2013)\citenamefont{Shcherbakov,
  Pe'er, Reynolds, Haas, Bode, and Laguna}}]{Shcherbakov:2012zt}
\bibinfo{author}{\bibfnamefont{R.~V.} \bibnamefont{Shcherbakov}},
  \bibinfo{author}{\bibfnamefont{A.}~\bibnamefont{Pe'er}},
  \bibinfo{author}{\bibfnamefont{C.~S.} \bibnamefont{Reynolds}},
  \bibinfo{author}{\bibfnamefont{R.}~\bibnamefont{Haas}},
  \bibinfo{author}{\bibfnamefont{T.}~\bibnamefont{Bode}}, \bibnamefont{and}
  \bibinfo{author}{\bibfnamefont{P.}~\bibnamefont{Laguna}},
  \bibinfo{journal}{Astrophys. J.} \textbf{\bibinfo{volume}{769}},
  \bibinfo{pages}{85} (\bibinfo{year}{2013}), \eprint{1212.4837}.

\bibitem[{\citenamefont{Ioka et~al.}(2016)\citenamefont{Ioka, Hotokezaka, and
  Piran}}]{Ioka:2016yzc}
\bibinfo{author}{\bibfnamefont{K.}~\bibnamefont{Ioka}},
  \bibinfo{author}{\bibfnamefont{K.}~\bibnamefont{Hotokezaka}},
  \bibnamefont{and} \bibinfo{author}{\bibfnamefont{T.}~\bibnamefont{Piran}},
  \bibinfo{journal}{Astrophys. J.} \textbf{\bibinfo{volume}{833}},
  \bibinfo{pages}{110} (\bibinfo{year}{2016}), \eprint{1608.02938}.

\bibitem[{\citenamefont{Greiner et~al.}(2015)}]{Greiner:2015lia}
\bibinfo{author}{\bibfnamefont{J.}~\bibnamefont{Greiner}} \bibnamefont{et~al.},
  \bibinfo{journal}{Nature} \textbf{\bibinfo{volume}{523}},
  \bibinfo{pages}{189} (\bibinfo{year}{2015}), \eprint{1509.03279}.

\bibitem[{\citenamefont{Levan et~al.}(2013)}]{Levan:2013gcz}
\bibinfo{author}{\bibfnamefont{A.~J.} \bibnamefont{Levan}}
  \bibnamefont{et~al.}, \bibinfo{journal}{Astrophys. J.}
  \textbf{\bibinfo{volume}{781}}, \bibinfo{pages}{13} (\bibinfo{year}{2013}),
  \eprint{1302.2352}.

\bibitem[{\citenamefont{MacLeod et~al.}(2014)\citenamefont{MacLeod, Goldstein,
  Ramirez-Ruiz, Guillochon, and Samsing}}]{MacLeod:2014mha}
\bibinfo{author}{\bibfnamefont{M.}~\bibnamefont{MacLeod}},
  \bibinfo{author}{\bibfnamefont{J.}~\bibnamefont{Goldstein}},
  \bibinfo{author}{\bibfnamefont{E.}~\bibnamefont{Ramirez-Ruiz}},
  \bibinfo{author}{\bibfnamefont{J.}~\bibnamefont{Guillochon}},
  \bibnamefont{and} \bibinfo{author}{\bibfnamefont{J.}~\bibnamefont{Samsing}},
  \bibinfo{journal}{Astrophys. J.} \textbf{\bibinfo{volume}{794}},
  \bibinfo{pages}{9} (\bibinfo{year}{2014}), \eprint{1405.1426}.

\bibitem[{\citenamefont{Abbott
  et~al.}(2017{\natexlab{a}})}]{TheLIGOScientific:2017qsa}
\bibinfo{author}{\bibfnamefont{B.~P.} \bibnamefont{Abbott}}
  \bibnamefont{et~al.} (\bibinfo{collaboration}{Virgo, LIGO Scientific}),
  \bibinfo{journal}{Phys. Rev. Lett.} \textbf{\bibinfo{volume}{119}},
  \bibinfo{pages}{161101} (\bibinfo{year}{2017}{\natexlab{a}}),
  \eprint{1710.05832}.

\bibitem[{\citenamefont{Abbott et~al.}(2017{\natexlab{b}})}]{Monitor:2017mdv}
\bibinfo{author}{\bibfnamefont{B.~P.} \bibnamefont{Abbott}}
  \bibnamefont{et~al.} (\bibinfo{collaboration}{Virgo, Fermi-GBM, INTEGRAL,
  LIGO Scientific}), \bibinfo{journal}{Astrophys. J.}
  \textbf{\bibinfo{volume}{848}}, \bibinfo{pages}{L13}
  (\bibinfo{year}{2017}{\natexlab{b}}), \eprint{1710.05834}.

\bibitem[{\citenamefont{Taylor et~al.}(2015)\citenamefont{Taylor, Ahlers, and
  Hooper}}]{Taylor:2015rla}
\bibinfo{author}{\bibfnamefont{A.~M.} \bibnamefont{Taylor}},
  \bibinfo{author}{\bibfnamefont{M.}~\bibnamefont{Ahlers}}, \bibnamefont{and}
  \bibinfo{author}{\bibfnamefont{D.}~\bibnamefont{Hooper}},
  \bibinfo{journal}{Phys. Rev.} \textbf{\bibinfo{volume}{D92}},
  \bibinfo{pages}{063011} (\bibinfo{year}{2015}), \eprint{1505.06090}.

\bibitem[{\citenamefont{Mucke et~al.}(2000)\citenamefont{Mucke, Engel, Rachen,
  Protheroe, and Stanev}}]{Mucke:1999yb}
\bibinfo{author}{\bibfnamefont{A.}~\bibnamefont{Mucke}},
  \bibinfo{author}{\bibfnamefont{R.}~\bibnamefont{Engel}},
  \bibinfo{author}{\bibfnamefont{J.}~\bibnamefont{Rachen}},
  \bibinfo{author}{\bibfnamefont{R.}~\bibnamefont{Protheroe}},
  \bibnamefont{and} \bibinfo{author}{\bibfnamefont{T.}~\bibnamefont{Stanev}},
  \bibinfo{journal}{Comput. Phys. Commun.} \textbf{\bibinfo{volume}{124}},
  \bibinfo{pages}{290} (\bibinfo{year}{2000}), \eprint{astro-ph/9903478}.

\bibitem[{\citenamefont{Koning et~al.}(2007)\citenamefont{Koning, Hilaire, and
  Duijvestijn}}]{Koning:2007}
\bibinfo{author}{\bibfnamefont{A.~J.} \bibnamefont{Koning}},
  \bibinfo{author}{\bibfnamefont{S.}~\bibnamefont{Hilaire}}, \bibnamefont{and}
  \bibinfo{author}{\bibfnamefont{M.~C.} \bibnamefont{Duijvestijn}}, in
  \emph{\bibinfo{booktitle}{{Proceedings, International Conference on Nuclear
  Data for Science and Tecnology}}} (\bibinfo{year}{2007}), pp.
  \bibinfo{pages}{211--214}.

\bibitem[{\citenamefont{Kampert et~al.}(2013)\citenamefont{Kampert, Kulbartz,
  Maccione, Nierstenhoefer, Schiffer, Sigl, and van Vliet}}]{Kampert:2012fi}
\bibinfo{author}{\bibfnamefont{K.-H.} \bibnamefont{Kampert}},
  \bibinfo{author}{\bibfnamefont{J.}~\bibnamefont{Kulbartz}},
  \bibinfo{author}{\bibfnamefont{L.}~\bibnamefont{Maccione}},
  \bibinfo{author}{\bibfnamefont{N.}~\bibnamefont{Nierstenhoefer}},
  \bibinfo{author}{\bibfnamefont{P.}~\bibnamefont{Schiffer}},
  \bibinfo{author}{\bibfnamefont{G.}~\bibnamefont{Sigl}}, \bibnamefont{and}
  \bibinfo{author}{\bibfnamefont{A.~R.} \bibnamefont{van Vliet}},
  \bibinfo{journal}{Astropart. Phys.} \textbf{\bibinfo{volume}{42}},
  \bibinfo{pages}{41} (\bibinfo{year}{2013}), \eprint{1206.3132}.

\bibitem[{\citenamefont{Puget et~al.}(1976)\citenamefont{Puget, Stecker, and
  Bredekamp}}]{Puget:1976nz}
\bibinfo{author}{\bibfnamefont{J.~L.} \bibnamefont{Puget}},
  \bibinfo{author}{\bibfnamefont{F.~W.} \bibnamefont{Stecker}},
  \bibnamefont{and} \bibinfo{author}{\bibfnamefont{J.~H.}
  \bibnamefont{Bredekamp}}, \bibinfo{journal}{Astrophys. J.}
  \textbf{\bibinfo{volume}{205}}, \bibinfo{pages}{638} (\bibinfo{year}{1976}).

\bibitem[{\citenamefont{Gilmore et~al.}(2012)\citenamefont{Gilmore, Somerville,
  Primack, and Dominguez}}]{Gilmore:2011ks}
\bibinfo{author}{\bibfnamefont{R.~C.} \bibnamefont{Gilmore}},
  \bibinfo{author}{\bibfnamefont{R.~S.} \bibnamefont{Somerville}},
  \bibinfo{author}{\bibfnamefont{J.~R.} \bibnamefont{Primack}},
  \bibnamefont{and}
  \bibinfo{author}{\bibfnamefont{A.}~\bibnamefont{Dominguez}},
  \bibinfo{journal}{Mon. Not. Roy. Astron. Soc.}
  \textbf{\bibinfo{volume}{422}}, \bibinfo{pages}{3189} (\bibinfo{year}{2012}),
  \eprint{1104.0671}.

\bibitem[{\citenamefont{Dominguez et~al.}(2011)}]{Dominguez:2010bv}
\bibinfo{author}{\bibfnamefont{A.}~\bibnamefont{Dominguez}}
  \bibnamefont{et~al.}, \bibinfo{journal}{Mon. Not. Roy. Astron. Soc.}
  \textbf{\bibinfo{volume}{410}}, \bibinfo{pages}{2556} (\bibinfo{year}{2011}),
  \eprint{1007.1459}.

\bibitem[{\citenamefont{Fang et~al.}(2017)}]{Fang:2017mhl}
\bibinfo{author}{\bibfnamefont{K.}~\bibnamefont{Fang}} \bibnamefont{et~al.},
  \bibinfo{journal}{PoS} \textbf{\bibinfo{volume}{ICRC2017}},
  \bibinfo{pages}{996} (\bibinfo{year}{2017}), \eprint{1708.05128}.

\end{thebibliography}

\clearpage


\appendix

\renewcommand{\appendixname}{Supplementary material}

\section{Details of the methodology}
\label{app:methods}
\begin{figure*}[t!]
\centering
\includegraphics[width=0.45\textwidth]{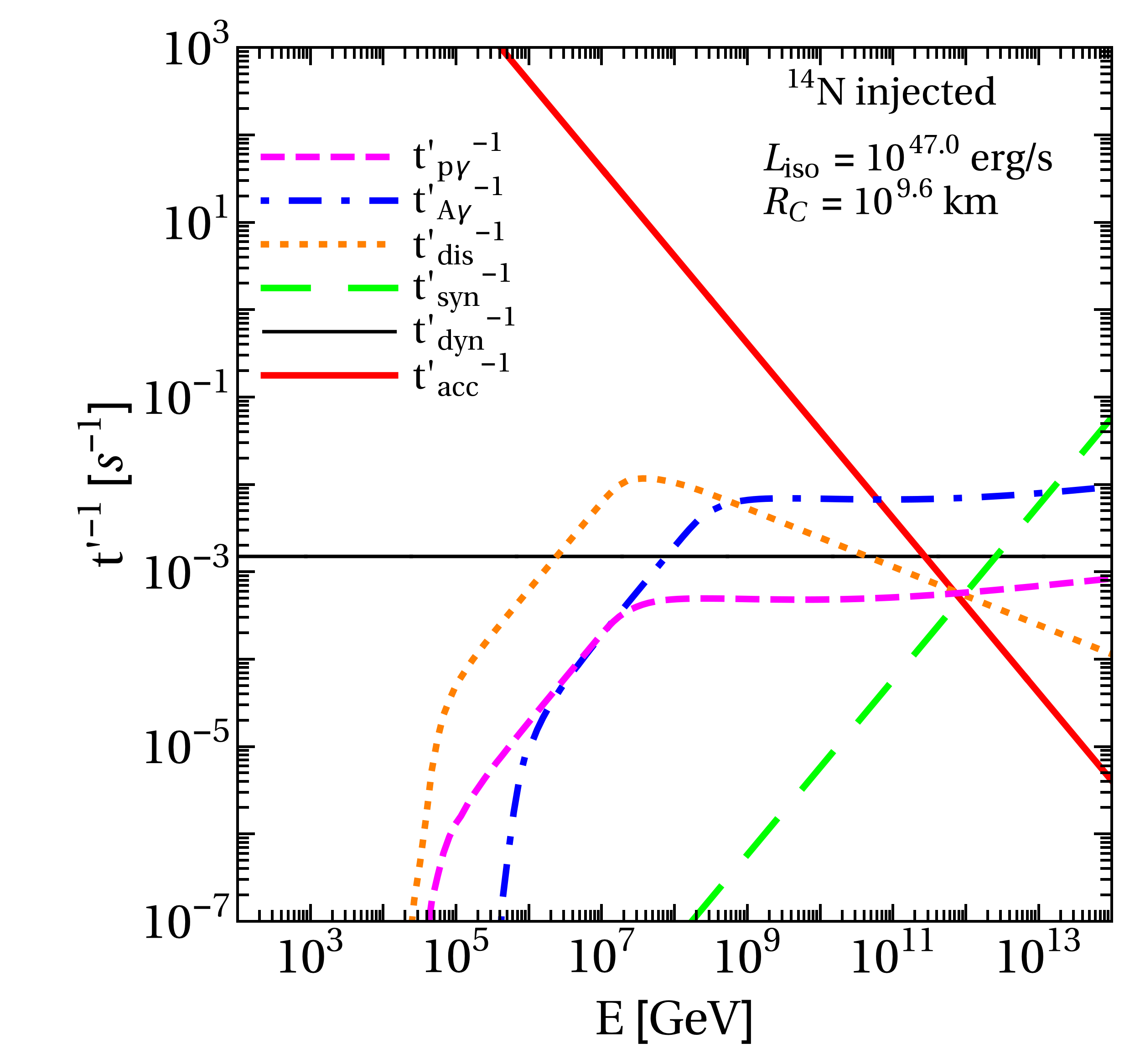}\hspace{0.5cm}
\includegraphics[width=0.45\textwidth]{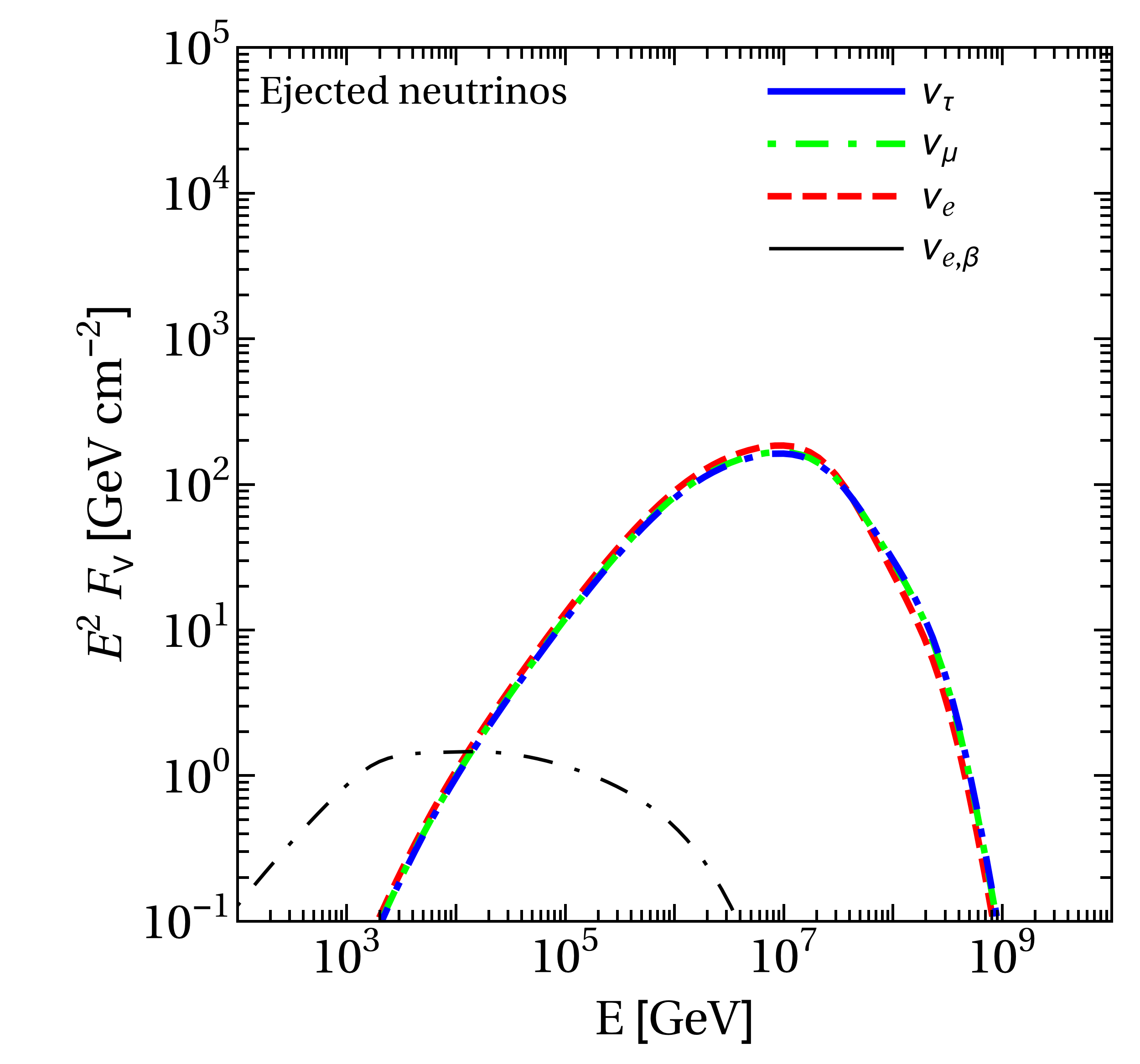}
\includegraphics[width=0.45\textwidth]{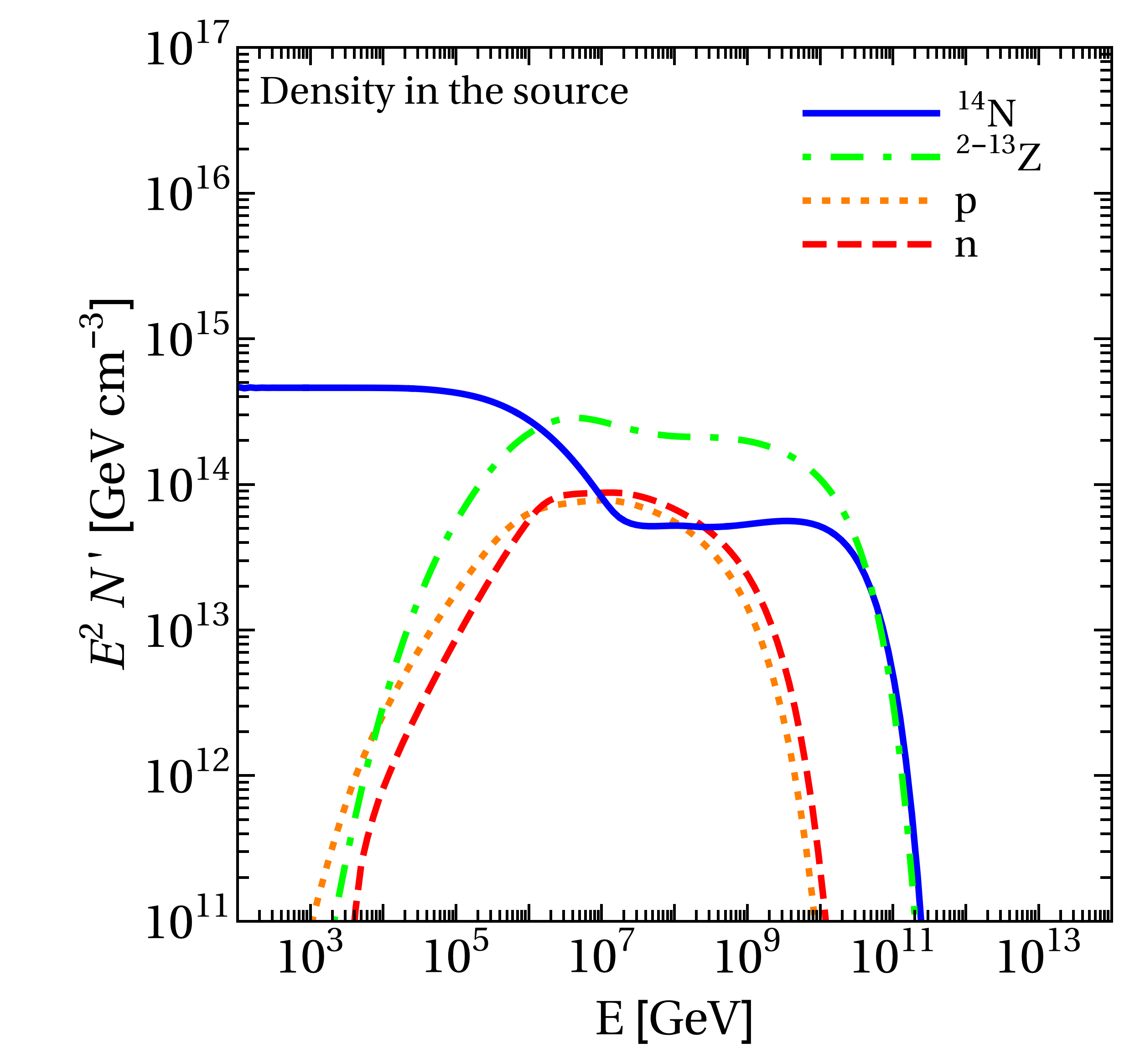}\hspace{0.5cm}
\includegraphics[width=0.45\textwidth]{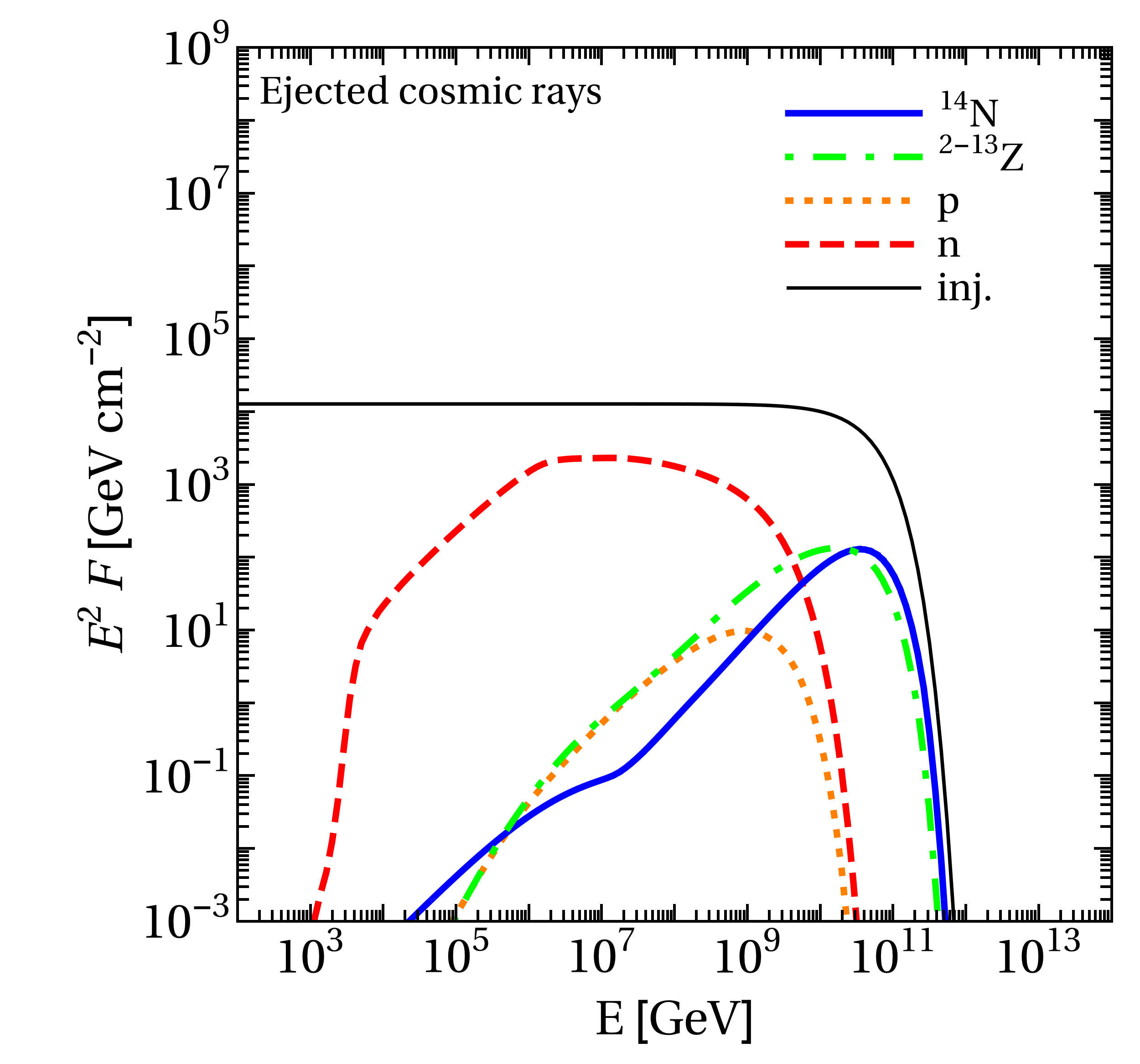}
\caption{Interaction rates (upper left), neutrino fluence per flavor (upper right), isotope density in the source (lower left) and ejected cosmic ray fluence (lower right, no interactions in the propagation included) as a function of the energy in the observer's frame at point~A in \figu{fit} ($L_X = 10^{47.0}$ erg s$^{-1}$ and $R = 10^{9.6}$ km) for pure $^{14}$N injection. The other TDE parameters are chosen to be $\Gamma = 10$, $\xi_A = 10$, $\varepsilon'_{\gamma,\text{br}} = 1$ keV and $z = 0.001$.}
\label{fig:proto}
\end{figure*}
\begin{figure}[t!]
\centering
\includegraphics[width=\columnwidth]{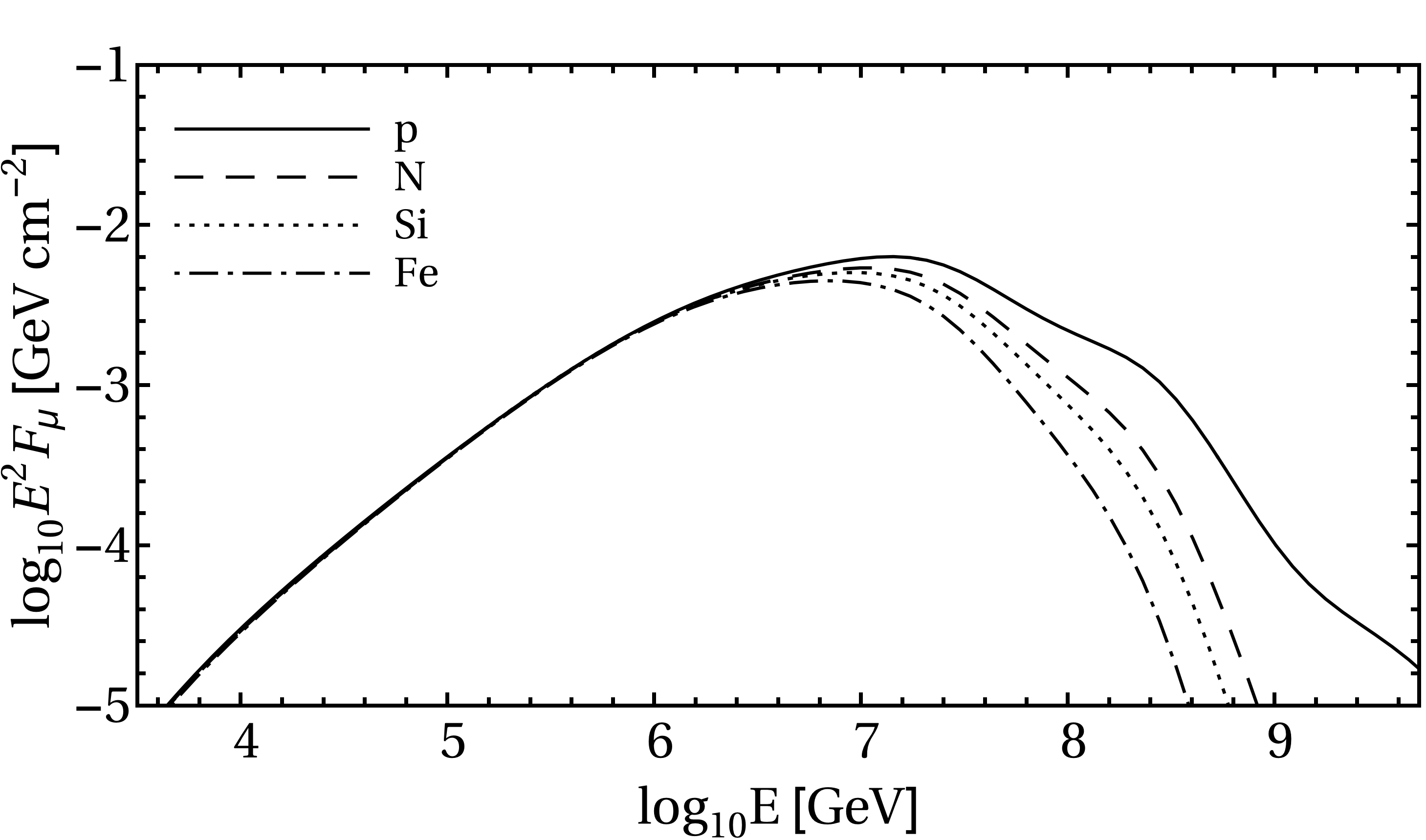}
\caption{Fluence of $\nu_\mu + \bar{\nu}_\mu$  for a single TDE  and different (pure) injection compositions. The chosen parameters are $z = 0.01$, $L_X = 47.5 \, \mathrm{erg \, s^{-1}}$, $R = 10^{9.8} \, \mathrm{km}$, and $\xi_A = 10$, as used in \cite{Lunardini:2016xwi} for the proton injection case (solid curve in this plot).\label{fig:nuflux}} 
\label{fig:nu-comp}
\end{figure}

Let us give details on the input parameters of the calculation. 
The target photon field is parameterized as a broken power law with a spectral break at $\varepsilon_{X,\text{br}} = 1$ keV in the observer's frame and spectral indices $\alpha = -2/3$ and $\beta = -2$ below and above the break energy, respectively. This choice is motivated by the spectral energy distribution (SED) of Swift J1644+57.
For this event, the isotropic equivalent luminosity of the X-ray flare was $L_X \simeq 10^{47.5}$ erg s$^{-1}$ over a time of $\Delta T \simeq 10^6$~s, leading to an estimated total energy of $E_X \simeq L_X \Delta T \simeq 3 \times 10^{53}$ erg. 
A Lorentz factor $\Gamma \sim 10$ and a minimum variability time $t_v \sim 10^2$~s are estimated from the observations \cite{Burrows:2011dn}. 
When performing the fit, we keep $\Gamma$ fixed and vary the collision radius $R$, therefore $t_v$ is determined by the relation $R = 2\Gamma^2 c t_v / (1+z)$ of the internal shock model.

The spectrum of the primary injected nuclei is taken of the form $\propto E^{-2} e^{-E/E_{\mathrm{max}}}$, where  $E_{\mathrm{max}}$ is obtained by balancing the acceleration with absorption or energy loss processes. 
The acceleration rate is  $t'^{-1}_\text{acc} = \eta c / R'_L$, where $R'_L = E'/ZeB'$ is the Larmor radius of a particle with charge number $Z$ and energy $E'$, and $\eta$  is the acceleration efficiency. The magnetic energy density is assumed to be in equipartition with the photon energy density.
We assume that acceleration is efficient, \ie, $\eta = 1.0$.  Note that $E_{\mathrm{max}}$ (and therefore, indirectly, $\eta$) is somewhat degenerate with the systematic shift of the measured UHECR energy (see main text). 

When simulating the nuclear cascade, we distinguish between two energy (photon energy in the nuclei's rest frame)  regimes, the  photo-disintegration ($\epsilon_\gamma \lesssim 150 \, \mathrm{MeV}$) and photo-meson production ($\epsilon_\gamma \gtrsim 150 \, \mathrm{MeV}$).
The photo-meson production simulation is based on the SOPHIA code~\cite{Mucke:1999yb}, with a superposition model for nuclei, \ie, $\sigma_{A\gamma} = A \sigma_{p\gamma}$ (with $A$ being the nucleus' mass number). The photo-disintegration uses the TALYS 1.8 code for nuclei with $A \geq 12$ \cite{Koning:2007} and CRPropa2 for lighter nuclei \cite{Kampert:2012fi} (for details see \cite{Boncioli:2016lkt}). 
\figu{proto} shows the interaction rates,  the neutrino fluence,  the isotope density in the source and the ejected cosmic ray fluence for the parameters of point~A in \figu{fit}. It appears that the maximum energy is limited by photo-meson production ($t'^{-1}_{A\gamma}$ exceeds $t'^{-1}_\text{acc}$ at $E_\text{max} \sim 6 \times 10^{10}$ GeV), implying that this is also the relevant process for disintegration at the highest energies.
 Note that in our superposition model for photo-meson production, which  still corresponds to the state-of-the-art in the literature, a nucleon is assumed to interact with the photon and then leaves the nucleus. The interaction of the single nucleon is described with SOPHIA, whereas the remaining nucleus is assumed to stay intact. A more  realistic model may involve additional disintegration of the remaining excited nucleus -- which is to be studied in the future. 
One can also see that the photo-disintegration rate follows the low-energy photon spectral index above the break at about $10^{7.5} \, \mathrm{GeV}$ (high-energy nuclei interact with low-energy photons), leading to a sub-dominant contribution at the highest energies (beyond about $10^{8.5} \, \mathrm{GeV}$) compared to the photo-meson production.

 \figu{proto} (upper right panel) shows the neutrino fluence in each flavor, in the observer's frame, computed for a source at redshift $z = 0.001$. Neutrinos from beta decays (from isotopes within and outside the source) are only relevant at low energies, and are shown as a separate curve.
 The plot of particle densities inside the source (lower left panel) shows that the nitrogen spectrum is depleted -- with respect to the $E^{-2}$ injection spectrum -- at the highest energies, where the isotopes produced in the disintegration chain dominate the spectrum.
The spectrum of the ejected neutrons (given in \figu{proto}, lower right panel) follows the spectrum  within the source (lower left panel).
Instead, the spectrum of the charged cosmic rays is harder (tilted in factor of higher energies), because we assume a direct \uh\ escape mechanism (for details see \cite{Baerwald:2013pu}). This mechanism conservatively assumes   that only particles from the boundaries of the production region within their Larmor radius can escape, and similar results are obtained for Bohm-like diffusion throughout the whole region. 
Here the escape is moderately efficient, as the Larmor radius is smaller than the size of the region at the highest energy (because the maximal energy is constrained by photo-hadronic interactions).

We also checked that the results in \figu{proto}  are consistent with those in \Ref~\cite{Zhang:2017hom}, when adjusted for the slightly different assumptions used there.  

In \figu{nuflux} we illustrate the dependence of the \n\ fluence of a single TDE on the (pure) injected nuclear  composition. Spectra for different nuclear species -- for the same injection luminosity --  are shown. The case of proton composition matches the corresponding one in \Ref~\cite{Lunardini:2016xwi}. It is evident that the change of composition affects the fluence mostly beyond its peak, at $E \gtrsim 10^7~{\rm GeV}$, as was already found in \cite{Biehl:2017zlw}.

In the current paper we discuss the C-O white dwarfs as possible origin of UHECRs and neutrinos, simulated by the most abundant isotope of nitrogen, $\mathrm{{}^{14}N}$ as representative for the most abundant isotope of carbon, $\mathrm{{}^{12}C}$, and oxygen, $\mathrm{{}^{16}O}$. Differences can be expected relative to a more realistic mixed C-O injection, both in the source and in the extragalactic propagation; some are due to the fact that for both $\mathrm{{}^{12}C}$ and $\mathrm{{}^{16}O}$, the $\alpha$-particle ejection is relevant, which could result in a slightly more efficient disintegration. The lack of cross section measurements for this channel \cite{Batista:2015mea,Boncioli:2016lkt} contributes to the uncertainties on the predictions of UHECR observables.

The propagation of the UHECRs between the source and Earth is modeled with the {\it SimProp} code \cite{Aloisio:2017iyh}, which takes into account nuclei photo-disintegration and photo-meson production, as well as the energy losses due to electron-positron pair production and to the redshift of energy. Simulations including the Puget-Stecker-Bredekamp (PSB) model \cite{Puget:1976nz} for photo-disintegration are used, while for the photo-meson production the cross section for single-pion production is employed. For the extragalactic background photon field, we follow the model in \cite{Gilmore:2011ks}. For UHECRs the horizon is limited to redshift $z\sim1$, however for the study of neutrinos produced during the propagation it was necessary to extend the simulations up to $z=6$.
A detailed discussion about the interface between the {\it NeuCosmA} code and the computation of the final observables through  {\it SimProp} is given in \cite{Biehl:2017zlw}.

The UHECR spectrum and composition measured by the Pierre Auger Observatory \cite{Valino:2015,Aab:2014kda} are fitted above $10^{19}$ eV, with the same technique used in \cite{Biehl:2017zlw}. A penalty for the overshooting of the flux at the lowest energies is included in the fit. 

\begin{figure*}[tp]
  \centering
  \includegraphics[width=0.45\textwidth]{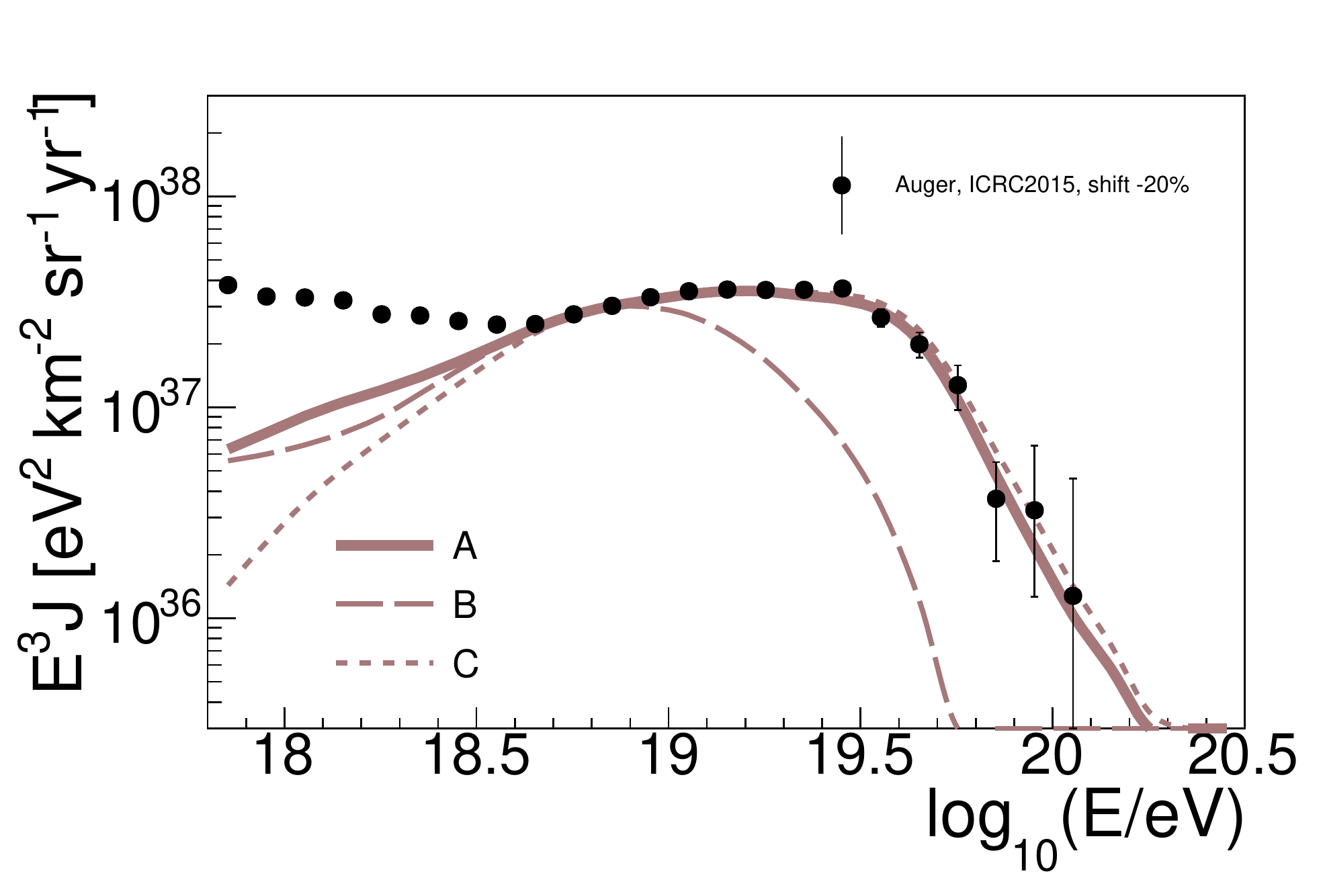}
  \includegraphics[width=0.45\textwidth]{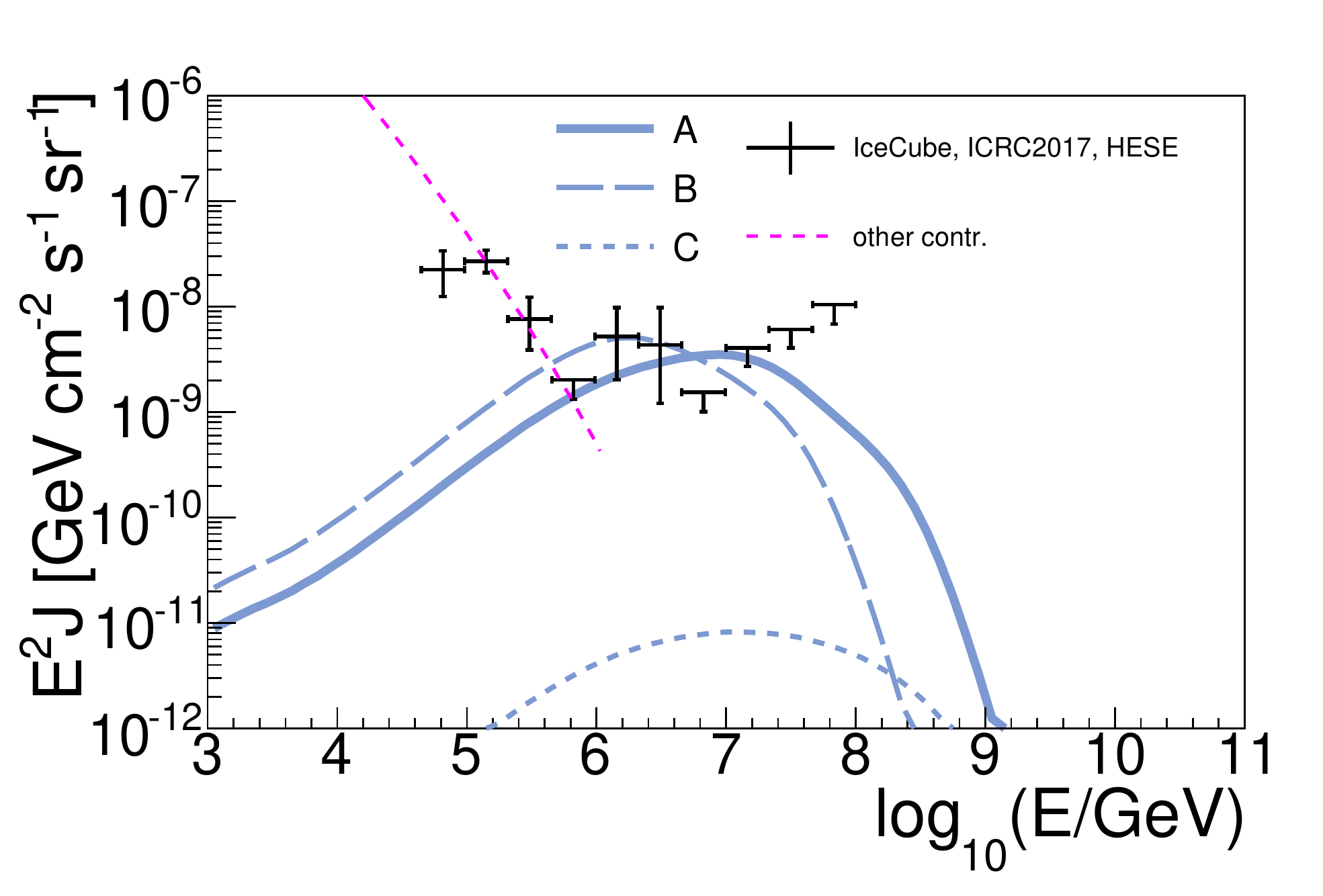}
  \\
\includegraphics[width=0.9\textwidth]{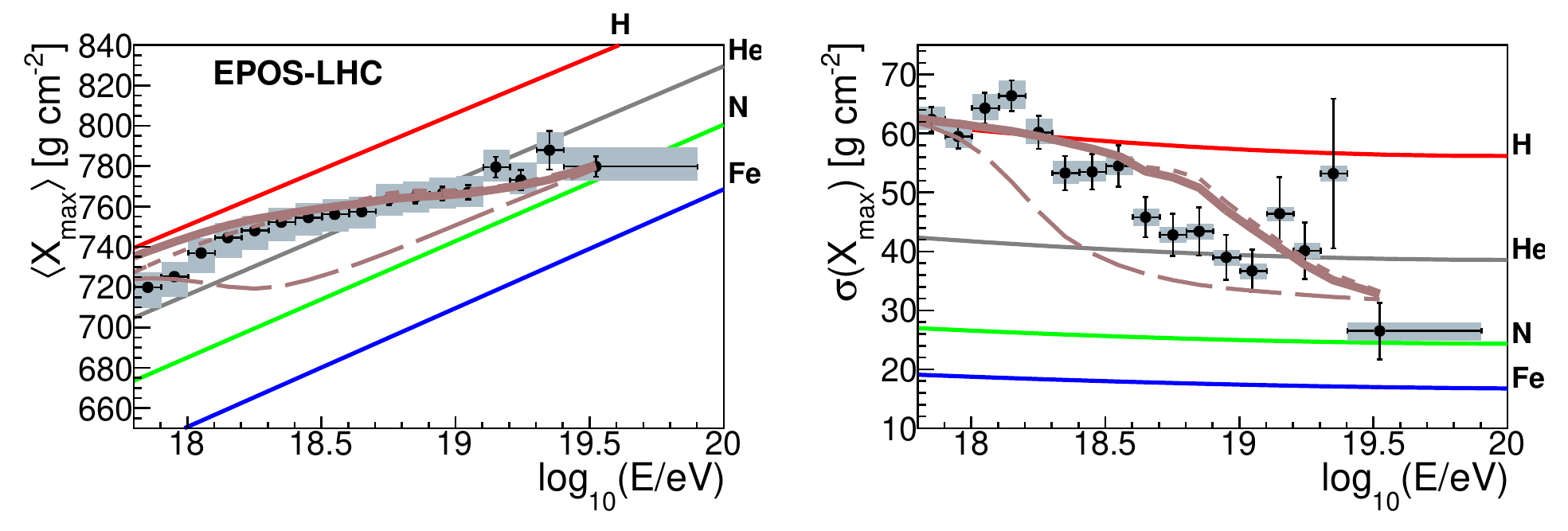}
\caption{Cosmic ray and (muon) neutrino observables obtained with the parameters corresponding to points A, B and C in \figu{fit}.}\label{fig:points}
\vspace{-0.3cm}
\end{figure*}

\begin{figure}[t!]
\centering
\includegraphics[width=\columnwidth]{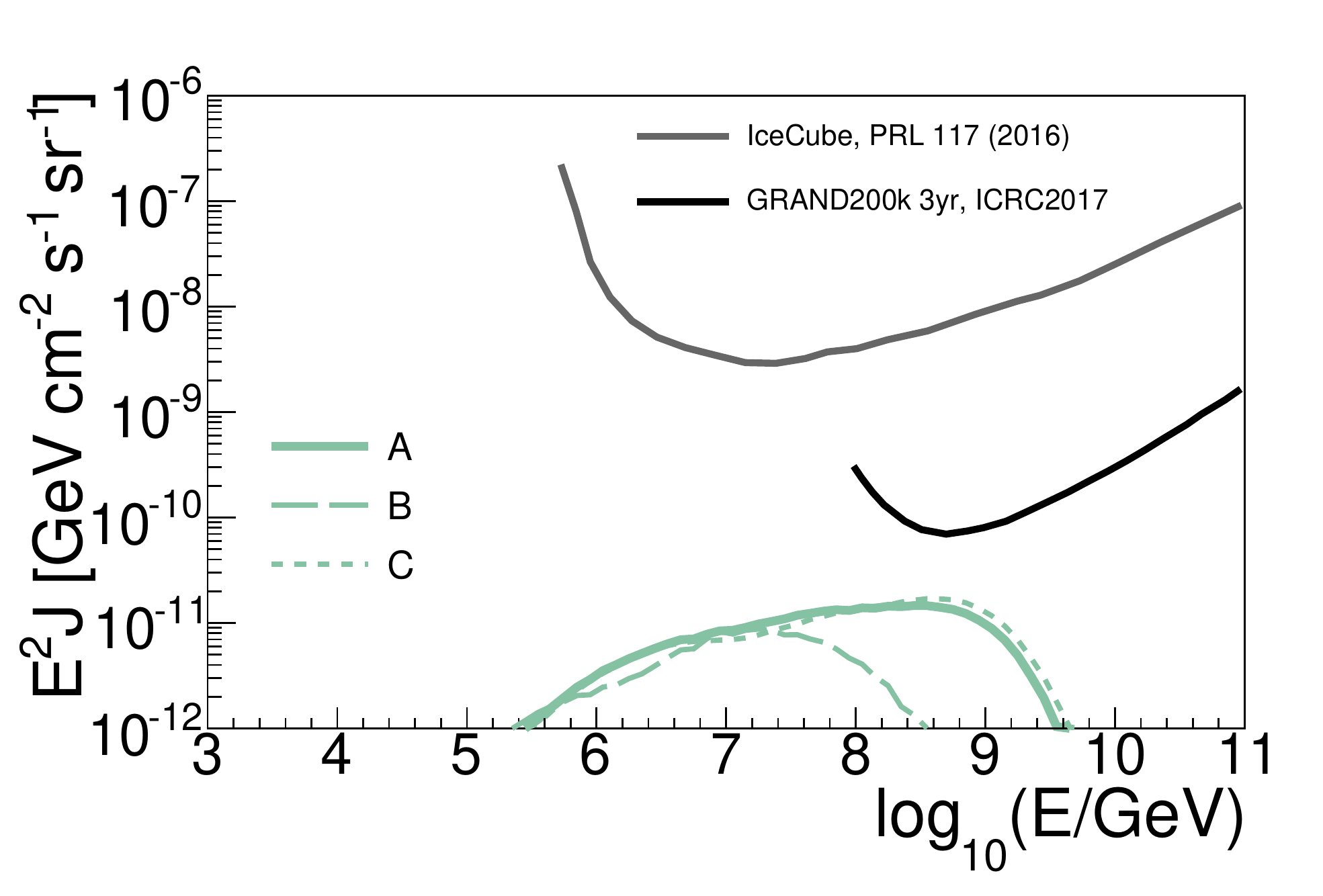}
\caption{Cosmogenic (muon) neutrino flux obtained with the parameters corresponding to points A, B and C in \figu{fit}.}\label{fig:points_cosmo}
\end{figure}

\section{Alternative fit scenarios and cosmogenic neutrinos}
\label{app:fit}

As a further illustration of the fit results, \figu{points} shows the same observables as in \figu{bestfit}, for the three parameter sets marked in \figu{fit} as A (same as in \figu{bestfit}), B (corresponding to the parameters which best describe the PeV neutrino data points), and C. It appears that for cases B and C one of the two data sets (\uh\ or \n) is underestimated, therefore a good joint description can not be found.

Let us discuss the effect of systematic uncertainties on the UHECR fit. 
To study the effect of uncertainties on \uh\ propagation, we generated results with alternative cross sections used in the photo-disintegration of nuclei, specifically the TALYS \cite{Koning:2007} cross sections for nuclei with $A\geq 12$ and cross sections as in \cite{Boncioli:2016lkt} for $A<12$. We find that the \uh\ best fit point corresponds to lower photon density, and is found within the $5\sigma$ region in \figu{fit}, which can therefore be considered as a robust  estimate of the allowed parameter space. 
A similar change in the position of the best fit is expected if an alternative model is used for the extragalactic background light, such as the one in \cite{Dominguez:2010bv}. It has a higher peak flux in the far infrared with respect to \cite{Gilmore:2011ks}, which implies  an increase of the photo-disintegration efficiency in a similar way as using a higher number of channels in the cross section model (as already pointed out in \cite{Batista:2015mea,Aab:2016zth}).

To account for the uncertain evolution with redshift of the \td\ rate, we repeated the calculations using a phenomenological rate, $\tilde R(z) \propto (1+z)^m$, with $m \geq 0$. Since the \uh\ spectrum is mostly sensitive to the local universe, it can be reasonably reproduced even with $m=0$ or 1, requiring a higher baryonic loading with respect to what found in the negative source evolution case. For $m\gtrsim 2$, the luminosity required to fit the \uh\ data results in overproducing the \n\ flux, and a good joint description of UHECRs and neutrinos is impossible. 

As an additional prediction of our model, we show the expected cosmogenic neutrino flux (for one flavor) in \figu{points_cosmo} for the same parameters as in \figu{points}.  The flux is well below the sensitivity of the GRAND experiment  \cite{Fang:2017mhl} (see figure). The suppression, compared to other predictions in the literature, is mostly due to the negative redshift evolution of the sources. An order of magnitude enhancement, reaching the design sensitivity of GRAND, is found for the alternative model with positive evolution, $m\simeq 2$. This implies, however, that the PeV neutrino data points are overshooted, as pointed out previously.

\section{Comparison to proton injection}
\label{app:proton}

\begin{figure*}[tp]
\includegraphics[width=0.53\textwidth]{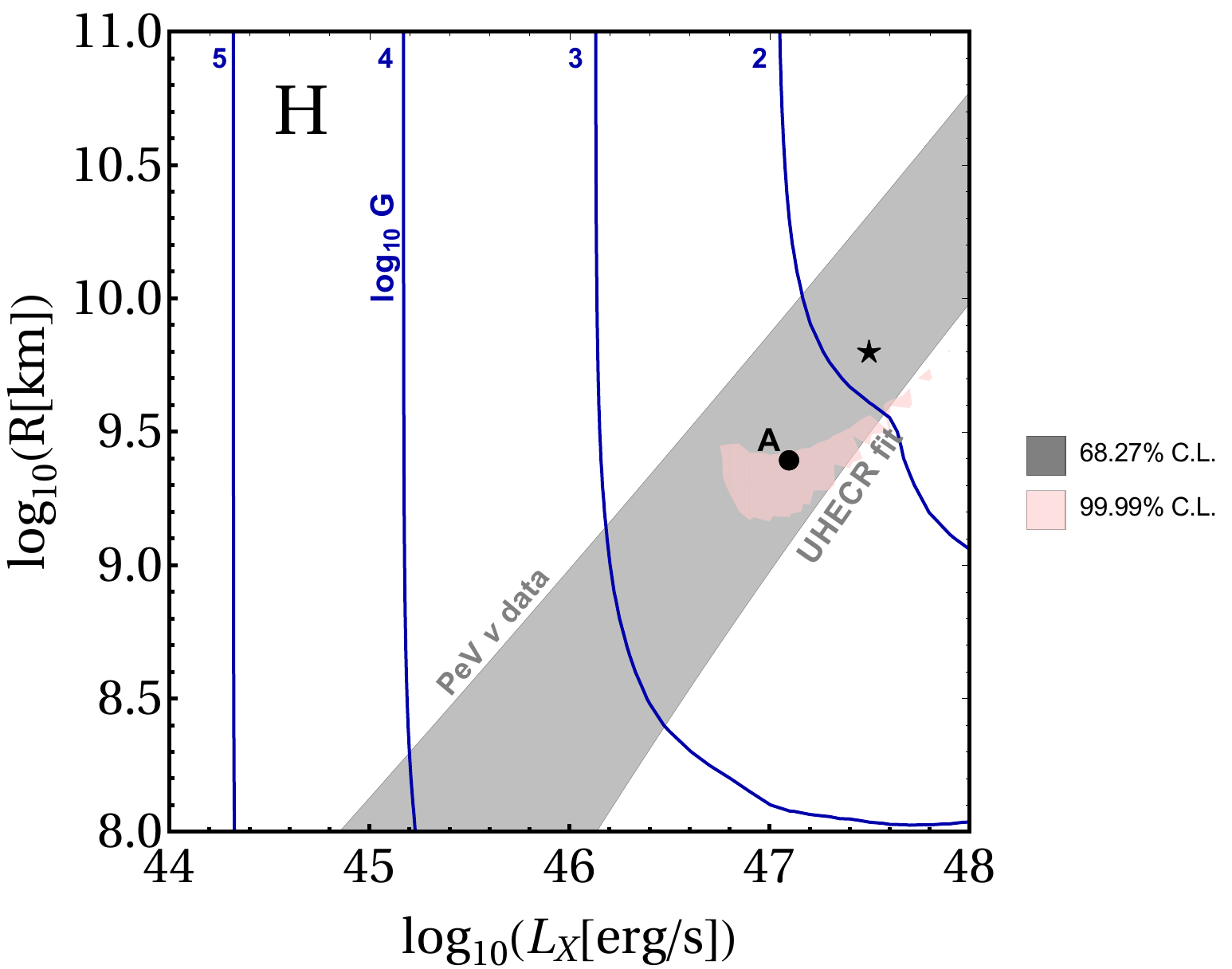} \hspace{0.4cm} 
\includegraphics[width=0.43\textwidth]{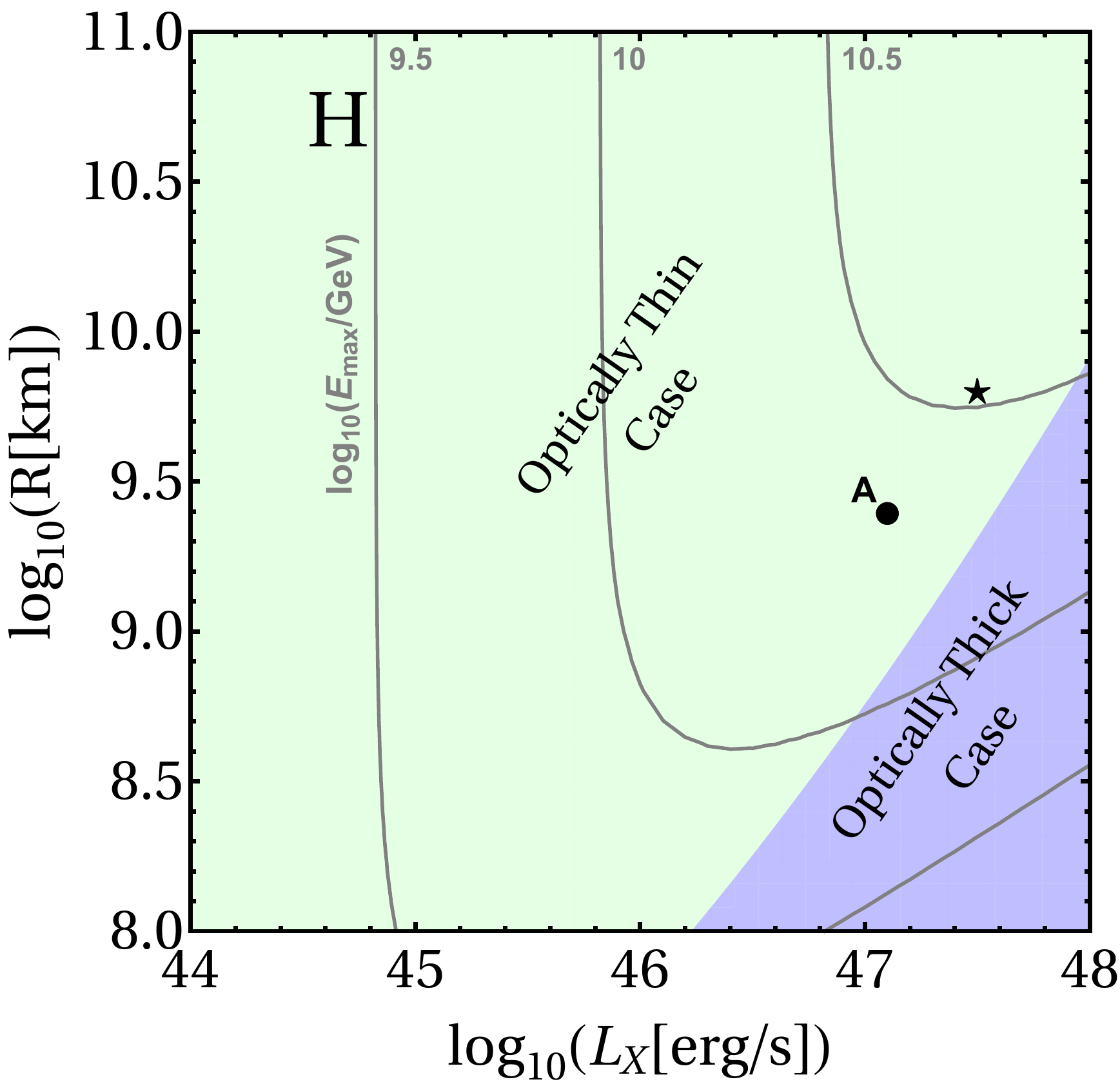}
\caption{Same as \figu{fit} in the main text, for the case of pure proton injection. Point A refers to the parameters which describes both UHECRs and PeV neutrinos, while the star refers to the standard case discussed in \cite{Lunardini:2016xwi} (base case).}\label{fig:fit_H}
\end{figure*}

\begin{figure*}[tp]
 \centering
 \includegraphics[width=0.45\textwidth]{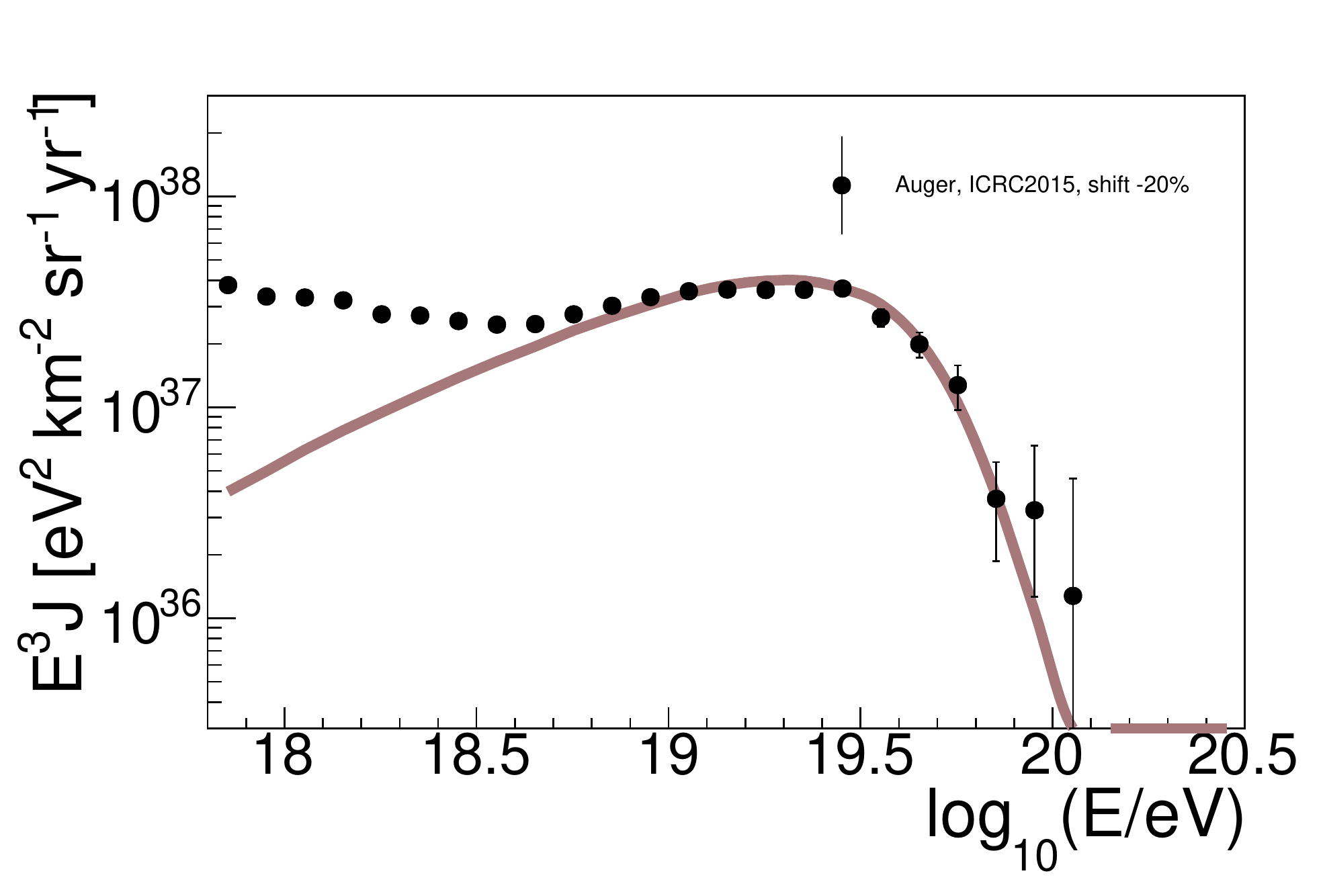}
 \includegraphics[width=0.45\textwidth]{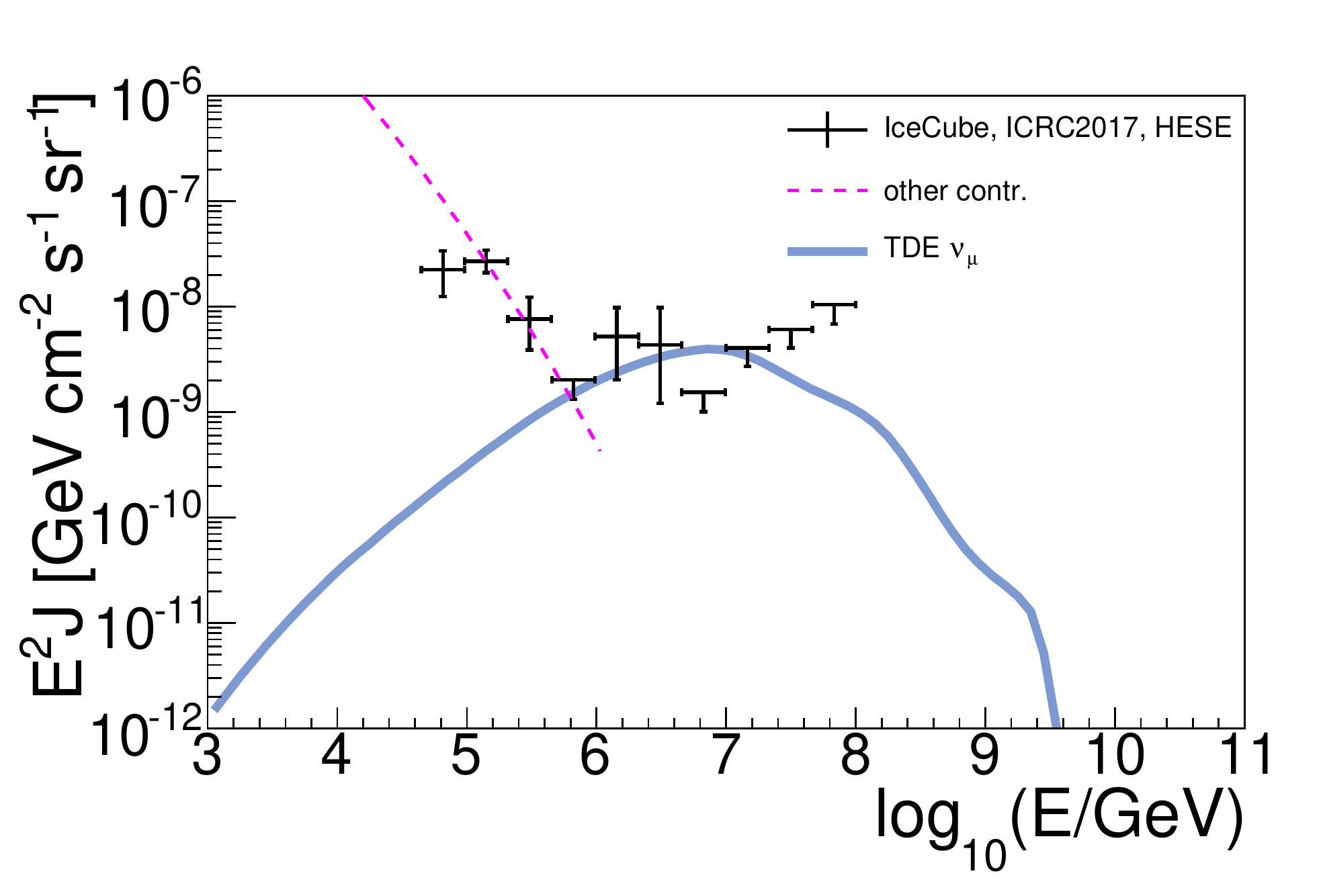}
\caption{Cosmic ray and (muon) neutrino fluxes obtained with the parameters corresponding to the combined fit (point A in \figu{fit_H}) with pure proton injection.}\label{fig:p_cr_prompt}
\vspace{-0.3cm}
\end{figure*}

\begin{figure}[t!]
\centering
\includegraphics[width=\columnwidth]{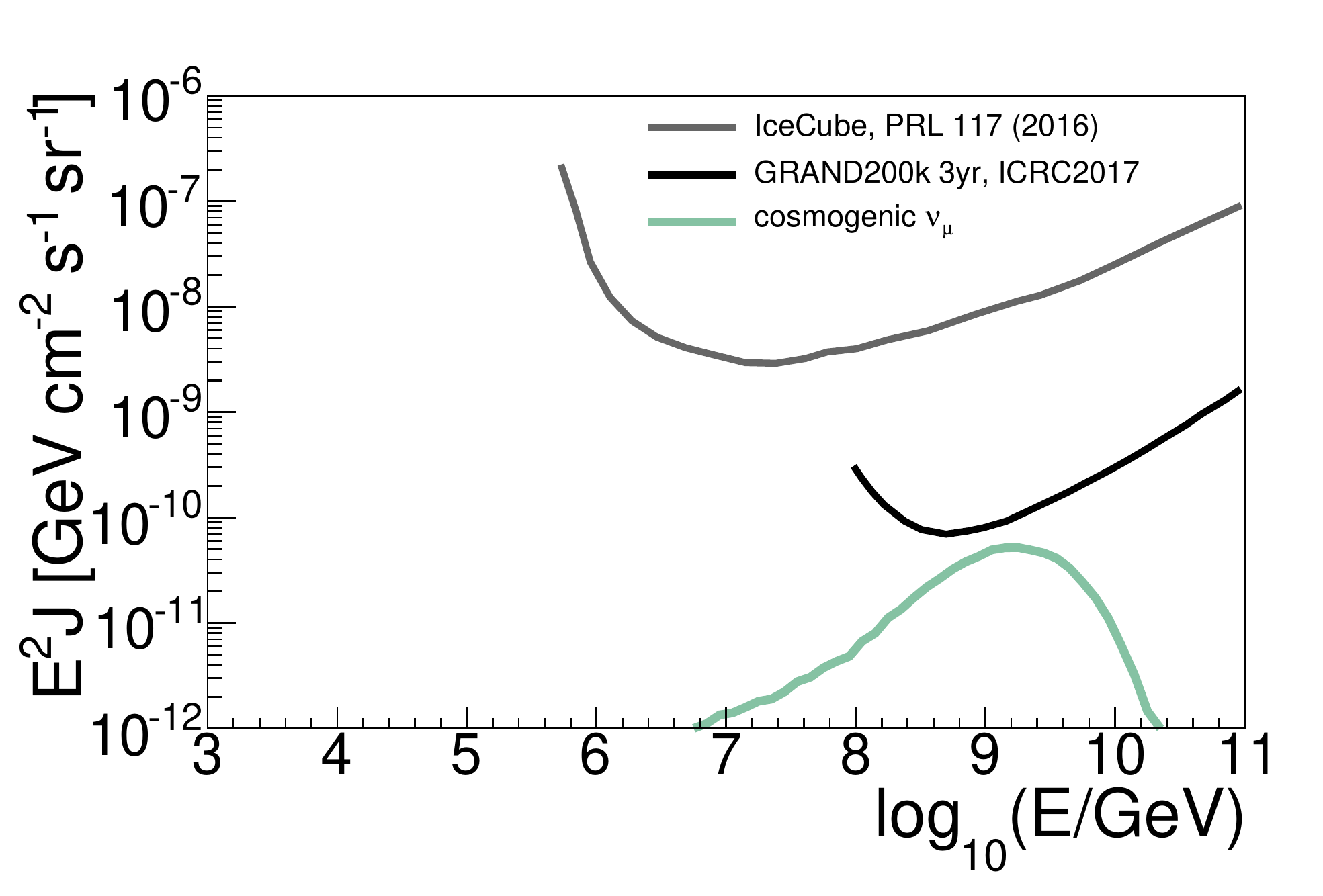}
\caption{Cosmogenic (muon) neutrino flux obtained with the parameters corresponding to the combined fit (point A in \figu{fit_H}) with pure proton injection.}\label{fig:p_cosmo}
\end{figure}

We show a set of plots here (\figu{fit_H}, \figu{p_cr_prompt}, and \figu{p_cosmo}) for the case of pure proton injection; while this scenario cannot describe the UHECR composition data, it may be interesting for comparison \eg\ to \Ref~\cite{Lunardini:2016xwi}. 

\Ref~\cite{Lunardini:2016xwi} demonstrates that the diffuse neutrino flux can be described using a pure proton composition in the TDE; the parameter choice corresponding to the base case scenario discussed in there is marked in \figu{fit_H} by a star. The proton-only fit corresponds to the optically thin (to photo-meson production) case, see \figu{fit_H}, right panel -- as it was indeed implied in \Ref~\cite{Lunardini:2016xwi}.
However, the  parameter choice in \Ref~\cite{Lunardini:2016xwi}
corresponds to a higher luminosity compared to  what has been found in the current study at the best-fit. As a consequence, the baryonic loading found in \cite{Lunardini:2016xwi} required  to power the diffuse neutrino flux  implies that the UHECR flux is too high by about a factor of seven (apart from different neutrino data used for reference there). This means that for protons, only about 10-15\% of the diffuse neutrino flux can be powered by jetted TDEs (in consistency with what has been found in \Refs~\cite{Dai:2016gtz,Senno:2016bso}). Since the required energy injection rate for UHECR nuclei is higher because of a shorter attenuation length, this problem does not occur for nuclei. 

A consequence of the proton composition at the source is the enhancement of the cosmogenic neutrino flux of a factor of $\sim 4$ with respect to the nitrogen case, as can be seen in \figu{p_cosmo}, resulting in a region within the design sensitivity of GRAND. However, such prediction is probably unrealistic, considering the poor fit of the pure proton scenario to the \uh\ data.

\end{document}